\documentclass[a4paper,fleqn,usenatbib]{mnras}

\usepackage{newtxtext,newtxmath}

\usepackage[T1]{fontenc}
\usepackage{ae,aecompl}

\usepackage{graphicx}	
\usepackage{amsmath}	
\usepackage{amssymb}	

\title[QPO frequency correlations in AMXPs]{On the frequency correlations of low-frequency QPOs with kilohertz QPOs in accreting millisecond X-ray pulsars}

\author[van Doesburgh and van der Klis]{
Marieke van Doesburgh,$^{1}$\thanks{E-mail: m.j.vandoesburgh@uva.nl}
Michiel van der Klis$^{1}$\\
$^{1}$Anton Pannekoek Institute, University of Amsterdam, Science Park 904, Postbus 94249, 1090 GE Amsterdam, The Netherlands\\
}
\date{Accepted XXX. Received YYY; in original form ZZZ}

\pubyear{2015}

\begin{document}
\label{firstpage}
\pagerange{\pageref{firstpage}--\pageref{lastpage}}
\maketitle

\begin{abstract}
We investigate frequency correlations of low frequency (LF, <80 Hz) and kHz quasi-periodic oscillations (QPOs) using the complete  RXTE data sets on 6 accreting millisecond X-ray pulsars (AMXPs) and compare them to those of non-pulsating neutron star low mass X-ray binaries with known spin. For the AMXPs SAX J1808.4--3658 and XTE J1807--294, we find frequency-correlation power law indices that, surprisingly, are significantly lower than in the non-pulsars, and consistent with the relativistic precession model (RPM) prediction of 2.0 appropriate to test-particle orbital and Lense-Thirring precession frequencies.  As previously reported, power law normalizations are significantly higher in these AMXPs than in the non-pulsating sources, leading to requirements on the neutron star specific moment of inertia in this model that cannot be satisfied with realistic equations of state. At least two other AMXPs show frequency correlations inconsistent with those of SAX J1808.4--3658 and XTE J1807--294, and possibly similar to those of the non-pulsating sources; for two AMXPs no conclusions could be drawn. We discuss these results in the context of a model that has had success in black hole (BH) systems involving a torus-like hot inner flow precessing due to (prograde) frame dragging, and a scenario in which additional (retrograde) magnetic and classical precession torques not present in BH systems are also considered. We show that a combination of these interpretations may accommodate our results. 
\end{abstract}

\begin{keywords}
X-rays: binaries -- accretion, accretion disks -- stars: neutron -- binaries: close
\end{keywords}



\section{Introduction}
Accreting low mass X-ray binaries (LMXBs) with a neutron star (NS) or black hole (BH) primary show remarkably similar timing behaviour. Typically, below 50 Hz, both their X-ray signals contain quasi-periodic oscillations (QPOs) with characteristic frequencies that shift as the source state changes, see \cite{Klis:2006book} for a review. 
At high frequencies (>100 Hz), however, NS show additional noise features and (twin) kHz QPOs with characteristic frequencies that correlate to those at low frequency.
In early work by \cite{vanStraaten:2003}, a match was presented between observations, and predictions by the relativistic precession model (RPM, \citealt{Stella:1998}) of frequency correlations between the upper kHz QPO frequency and a feature $<$50 Hz in three non-pulsating atoll NS-LMXBs.
In the RPM, the kHz QPO occurs at frequency $\nu_K$ of an orbit, and a $\sim$20 Hz LF-QPO is the Lense-Thirring (LT) precession frequency ($\nu_{LT}$) of that orbit due to frame dragging:
\begin{align}
 \nu_{LT}
    &= \frac{ 8\pi^{2}\emph{I}\nu_{K}^{2}\nu_{s}}{\emph{c}^{2}\emph{M}} = 13.2\text{ }\emph{I}_{45}\emph{m}^{-1}\nu_{\emph{K,3}}^{2} \nu_{\emph{s,2.5}}\text{ Hz }, 
    \label{eq:LT}
\end{align} where $\nu_{\emph{s}}=300\nu_{s,2.5}$ Hz is the neutron star spin frequency, $\nu_{\emph{K}}=10^3\nu_{K,3}$ Hz the Keplerian orbital frequency. $M=m$$\cdot$$M_{\odot}$ and $I=10^{45}I_{45}$ g cm$^{2}$ are the neutron star mass and moment of inertia, respectively.
For realistic equations of state $I_{45}/m$ ranges from 0.5 to 2 \citep{Friedman:1986}. The frequency of the low frequency QPO is predicted to be proportional to the spin frequency and quadratically related to the upper kHz QPO frequency in the same way as for a precessing test-particle orbit. Remarkably,  \cite{vanStraaten:2003} found power-law indices for correlations of the kHz QPO frequency with a LF feature of 2.0$\pm$0.02.

In \cite{vandoesburgh:2016} (hereafter DK17) we performed a deeper investigation of this result focusing on the non-pulsating sources with known spin. As the RPM predictions are for fundamentally periodic phenomena made quasi-periodic by 'clump' lifetime broadening and moderate frequency modulation associated with orbital decay \citep{Stella:1998, Lamb:2003} we directly measured the QPO centroid frequencies ($\nu_0$) expected to be representative of these periodicities instead of deriving them from the so-called characteristic frequencies ($\nu_{max}$, \citealt{Belloni:2002}). 
Due to a more careful analysis involving less data averaging, using bigger data sets, and including more sources, we were able for the first time to distinguish between two different low-frequency components which had been confused in earlier works. Both of these were found to have (centroid) frequency correlations with the kHz QPO frequencies with power law indices significantly larger than 2.0, in conflict with the RPM prediction. While the best-fit power laws to the frequency correlations differed from each other significantly in both index and normalization, neither correlated to spin in any obvious way. We concluded that LT-precession of near-test-particle orbits can not explain the QPOs. \\
Recent advances in the combined energy spectral and timing analysis of BH systems (spectral-timing of LF-QPOs) show that a misaligned toroidal hot inner flow precessing like a solid body around the BH spin-axis due to frame-dragging is a strong candidate-explanation for the source behaviour \citep{Ingram:2016}. 
As we noted in DK17 this physical picture could also explain the indices exceeding 2.0 for frequency correlations in non-pulsating NS-LMXBS: in the hot flow model of \cite{Ingram:2009} and \cite{Ingram:2010} the kHz QPO is associated with the orbital frequency at the inner edge of the outer thin accretion disk at radius $r_o$, and the low frequency QPO is thought to come from the precessing hot inner flow that is situated inside $r_o$. Steeper indices on the frequency correlations than the test-particle case may occur because $r_o$ decreases so that the torus narrows as the kHz QPO frequency becomes higher. 
We note that comparisons of LF QPOs in NS and BH systems have revealed parallels that hint at a common origin (see i.e. \citealt{Psaltis:1999}). However, due to the scarce detections of HF QPOs in BH systems, commonality between high frequency QPOs in BH and NS systems has not been convincingly established \citep{Belloni:2012}. \\
Other models that may explain some aspects of the QPOs and their frequency correlations include resonant disk-oscillation mode (discoseismic) models (i.e. \citealt{Kato:2001, Kato:2007}). Oscillations due to resonance of orbital frequencies in geometrically thick tori have also been explored, but most of these efforts focus on accommodating observations of high frequency QPOs only (i.e. \citealt{Rezzolla:2003, Abramowicz:2006}).\\
Another possible explanation of the steep frequency correlations  combines LT, magnetic, and classical precession around the NS to produce a net precession frequency whose correlation with the orbital frequency depends on the interplay between the torques \citep{Shirakawa:2002}. 
If for example the magnetic torque dominates, a steeper power law may be obtained. Further insight in some of these issues might be gained by studying the accreting millisecond X-ray pulsars (AMXPs), as these sources are likely to have a different magnetic field configuration than the non-pulsars. 
We excluded the AMXPs from DK17, because work by \cite{vanStraaten:2005} and \cite{Linares:2005} had shown that the frequency correlations of the AMXPs SAX J1808--3658, XTE\ J1807--294 and  XTE\ J0929--314 were systematically shifted towards lower $\nu_u $ (by a factor $\sim$1.5) compared to those of the non-pulsating atoll sources 4U\ 0614+09, 4U\ 1608-52, 4U\ 1728--34 and Aql X-1.
The non-pulsating NS-LMXB 4U 1820-30, whose spin is unknown, was reported to show frequency correlations shifted  similarly to those of the AMXPs mentioned \citep{Altamirano:2005}. 
An aperiodic timing analysis of the complete RXTE data set on SAX\ J1808--3658, confirming the shift, was later presented by \cite{Bult:2015b}.
In all these works characteristic frequencies ($\nu_{max}$) were used (sometimes converted to centroid ($\nu_0$) ones), so those reported best-fit power law indices are not directly comparable to those we reported in DK17. 
In the current work, to obtain a complete overview of centroid frequency correlations for all sources with known spin, we present the analysis of 4 AMXPs that show the required (L$_u$, L$_{LF}$) QPOs. We add two relevant non-pulsating atoll sources: 4U 1820-30 for reasons mentioned above, and SAX J1810.8--2609 which was recently added to the list of known-spin sources through the detection of 531.8$\pm{0.2}$ Hz burst oscillations by \cite{Bilous:2018}. 
For all these sources we present the best-fit power law indices and normalizations and their joint confidence contours, which allows for a straightforward comparison of the frequency correlations found in the different source types. For SAX\ J1810.8--2609 our work additionally constitutes the discovery of kHz and LF QPOs.
\\\\

\section{Methods, Observations}
Our source sample consists of the 6 AMXPs observed with RXTE known to have kHz QPOs: SAX J1808.4--3658,  XTE J1807--294, HETE J1900.1--2455, IGR J17480--2446, SAX J1748.9--2807, and IGR J17511--3057 (hereafter SAX J1808, XTE J1807, HETE J1900, IGR J17480, SAX J1748 and IGR J17511, respectively) and additionally 4U 1820--30 and SAX\ J1810.8--2609 (hereafter 4U 1820 and SAX\ J1810).  
In \cite{vanStraaten:2005}, characteristic frequencies are also reported for the AMXPs XTE J1751-305, XTE J1814-338 and XTE J0929-314. Most of the RXTE data were taken when these sources were in the island state (high hard colour, see Section \ref{sec:spec}), where the upper kHz QPO is broad and the centroid frequency of the fitted Lorentzian is ill defined. For this reason we exclude these sources from our sample. 

 \subsection{Data sets}
If available, we rely on earlier timing studies to find the data suitable to our analysis. If no such study exists, or if more data have become available since, we select the data with simultaneous LF- and kHz QPOs by visual inspection of power spectra of appropriate resolution and integration time. All selected RXTE data are listed in Table \ref{tab:pars2}. 

\subsection{Spectral analysis}
\label{sec:spec}
 Following the method described in \cite{vanStraaten:2003}, we calculate Crab-normalized hard and soft colors using Standard 2 data to track the accretion state of the source. The soft colour is defined as the count rate in the 3.5-6.0 keV band divided by the count rate in the 2.0-3.5 keV band. The hard colour uses the count rate in the 9.7-16.0 band divided by the count rate in the 6.0-9.73 keV band. We generally avoid averaging observations that differ in colour by more than 10$\%$ or that were taken more than a day apart. In some cases however, observations were grouped as reported in earlier work according to different criteria. These groups are explicitly mentioned in Appendix \ref{appendix:a}.
 
\subsection{Timing analysis}
We use archival RXTE data recorded in Event, Binned or Good Xenon mode with a time resolution of at least 1/8192 s ($\sim$122 $\mu$s).\\
We rely on previously reported best energy bands for QPO detection, that can differ per source. If nothing is specified in the literature we use a 2-20 keV energy band, where a high signal-to-noise is expected. We always confirm per source that the 2-20 keV energy band increases the detection significance of LF and kHz QPOs compared to using all energy channels. Unless noted otherwise, we calculate FFTs using 16-second data segments. \\
 After averaging the Leahy-normalized power spectra, we correct for background and deadtime effects by subtracting a counting noise model spectrum \citep{Zhang:1995} following the method of \cite{Klein:2004PhD} and renormalizing the power spectra such that the square root of the integrated power in the spectrum equals the fractional root mean square (rms) of the variability in the source signal \citep{Klis:1989}. 

If one is present, we remove the pulsar spike before rebinning and fitting the power spectra. We fit a multi-Lorentzian model to characterize the power spectral features, where each Lorentzian is characterized with the parameters centroid frequency ($\nu_0$), full width at half maximum (FWHM) and fractional rms amplitude (rms).

We use the naming scheme presented in
\cite{Altamirano:2008} (based on \citealt{vanStraaten:2002}) where features are identified as break (L$_b$), second break (L$_{b_2}$), low frequency QPO (L$_{LF}$), harmonic of the low frequency QPO (L$_{LF_2}$), hump (L$_{h}$), hectoHz (L$_{hHz}$), low frequency Lorentzian (L$_{{\ell}ow}$), lower kHz QPO (L$_{\ell}$), or upper kHz QPO (L$_u$). This scheme relies on the location of components in the power spectrum,  the correlations between their characteristic frequencies ($\nu_{\max}$) and the similarities in the appearance of power spectra between different sources. 

We fit asymmetrical low-frequency (so-called "flaring" \citealt{Bult:2015}) features in the power spectra of 4U\ 1820--30 and SAX J1808 with multiple Lorentzians; we name these features L$_{F_n}$ with n=1,2 or 3. This approach, instead of for instance fitting a Schechter function \citep{Bult:2015b} does not significantly affect the centroid frequency of features at higher frequency ($>$10 Hz).  We do not use the frequencies of the L$_{F_n}$ for correlation fitting. 

We write $\nu_i$, FWHM$_i$ and rms$_i$ for the centroid frequency, FWHM and fractional rms of component L$_i$.\\
For the sources included in DK17, the broad L$_b$ noise feature evolves into a QPO toward high (>700 Hz) $\nu_u$. For the sources included in the current sample, the range of $\nu_u$ is limited (and typically <700 Hz), and at low frequency the asymmetric flaring features complicate identification and comparison of power spectral components between sources. For these reasons we do not present power law fits to ($\nu_u$,$\nu_b$) frequency pairs here, but focus instead on the ($\nu_u$, $\nu_{LF}$) and ($\nu_u$, $\nu_h$) correlations.\\
We use our new optimal extraction method to measure power law indices and normalizations and their errors directly from the ensemble of power spectra in an unbiased way. In DK17
a detailed discussion of this method can be found.
\section{Results}

In Table \ref{sources} we list the sources included in our sample, along with the total number of observations (each typically $\sim$2 ks in length) in the RXTE archive. We carefully examined these to identify the observations that simultaneously show the relevant QPOs and hence can be used for our analysis.
The power spectrum of each available archival observation was individually considered, and combinations of observations with similar colours to increase QPO significance were explored. 
Sources that have a low number of useful observations are typically weak, and/or are usually observed at low or high hard colour where LF and kHz QPOs are normally detected at low significance or not at all. For IGRJ 17511--3057 and SAX J1748.9--2807 no simultaneous kHz and LF-QPOs were detected.

\begin{table}
\centering
  \tabcolsep=0.1cm
 {
\scalebox{0.9}{
\small
	\begin{tabular*}{\columnwidth}{@{\extracolsep{\fill}}lccc}
   \hline 
\vspace*{0.05cm}
 Source  & Spin (Hz) & Obs. in Archive  & Obs. used  \\
\hline
SAX J1808.4--3658 & 401.0 & 514  & 46 \\
XTE J1807--294 & 190.6 & 116 & 12 \\
HETE J1900.1--2455 & 377.3 & 363 & 14 \\
IGR J17480--2446 & 11.0 & 117 & 3\\
SAX J1748.9--2807 & 442.4 & 174 & 0 \\
IGR J17511--3057 & 244.8 & 290 &  0  \\
4U\ 1820--30 & N.A. & 234 & 26 \\
SAX J1810.8--2609 & 531.8$\pm{0.2}$ & 105 & 2 \\
\hline 
  \end{tabular*}
  }}
  \caption{The sources included in our sample. The neutron star spin frequency was inferred either from (intermittent, in the case of HETE\ J1900.1--2455) pulsations \citep{Watts:2012, Ritter:2003} or from burst oscillations (in the case of SAX\ J1810.8--2609, \citealt{Bilous:2018}).}
  \label{sources}
\end{table}

\subsection{Identification of power spectral components}
A detailed overview of the identification of features for each source can be found in Appendix \ref{appendix:a}. In Table \ref{tab:pars2}, a selection of relevant fit components are listed; the complete table can be found as an online appendix to this paper.

\subsection{Fit results}
\begin{table*}
\centering
  \tabcolsep=0.1cm
 {
\scalebox{1}{
\small
  \begin{tabular}{l c c c c c c }

   \hline 
\vspace*{0.05cm}
 Source  & L$_{i}$ & Group & No. of freq. &  $N$ (Hz) & $\alpha$ & Reduced  \\
  & & &  pairs ($\nu_i$, $\nu_u$) & & &  $\chi^2$ ($dof$)  \\
\hline
SAX J1808.4--3658 & LF & 1 & 17 & 37.1$\pm{0.3}$  & 2.00$\pm{0.03}$ & 1.18 (5455)  \\[0.1cm]
 & h & 1 & 41 & 51.3$\pm{0.3}$  & 1.89$^{+0.03}_{-0.02}$ & 1.12 (13338) \\[0.1cm]
 & \textbf{LF} & \textbf{2} & \textbf{10} & \textbf{36}$\pm{\textbf{0.4}}$  & \textbf{2.01}$\pm{\textbf{0.04}}$ & \textbf{1.12 (3163)} \\[0.1cm]
 & \textbf{h} & \textbf{2} & \textbf{9} & \textbf{56.0}$\pm{\textbf{0.7}}$  & \textbf{2.01}$\pm{\textbf{0.04}}$ & \textbf{1.09 (2839)} \\[0.1cm]
 \hline
XTE J1807--294 & \textbf{LF} & \textbf{1} & \textbf{12} & \textbf{46.4}$\pm{\textbf{1.4}}$  & \textbf{2.11}$\pm{\textbf{0.08}}$ & \textbf{1.02 (3922)}  \\[0.1cm]
\hline
HETE\ J1900.1--2455 & LF & 1 & 3 & 15.8$\pm{0.9}$  & 2.1$\pm{0.4}$ & 0.97 (1328) \\[0.1cm]
& \textbf{h} & \textbf{1} & \textbf{8} & \textbf{29.7}$^{\textbf{+1.3}}_{\textbf{-0.5}}$  & \textbf{2.4}$\pm{\textbf{0.1}}$ & \textbf{1.02 (3420)} \\[0.1cm]
\hline
IGR\ J17480--2446 & LF & 1 & 3 & 19.6$\pm{0.4}$  & 2.4 ($\textit{fixed}$) & 1.4 (994)\\[0.1cm]
\hline
SAX\ J1810.8--2609 & LF & 1 & 2 & 20.5$\pm{2.0}$  & 2.0$\pm{0.3}$ & 0.95 (578) \\[0.1cm]
\hline
4U 1820--30 & LF & 1 & 18 & 22.2$\pm{0.35}$  & 2.11$^{+0.04}_{-0.03}$ & 1.02 (5858) \\[0.1cm]
 & \textbf{h}& \textbf{1} & \textbf{8} & \textbf{38.8}$\pm{\textbf{0.9}}$  & \textbf{2.6}$\pm{\textbf{0.2}}$ & \textbf{0.98 (2558)} \\[0.1cm]
 & \textbf{LF} & \textbf{2} & \textbf{8} & \textbf{22.3}$\pm{\textbf{0.4}}$  & \textbf{2.26}$^{\textbf{+0.12}}_{\textbf{-0.11}}$ & \textbf{0.98 (2558)} \\[0.1cm]

 \hline
 \end{tabular}
  }}
  \caption{Power law fit parameters for $\nu_i$ vs. $\nu_u$ correlations. $N$ is the power law normalization at $\nu_u$=600 Hz, $\alpha$ is the power law index. See the Appendix for explanation of groups. Errors quoted here use $\Delta\chi^2$=1. We indicate in \textbf{bold} the data selections with more secure power spectral identifications whose contours are plotted in Figures \ref{fig:LFcontours} and \ref{fig:Hcontours}.
  }
  \label{fit}
\end{table*}

In Table \ref{fit}, we list the best-fit index and normalization of power laws fitted to the correlated frequencies in different data groups for each source with sufficient data quality. Group 1 contains all frequencies we measured, group 2 consists of a subselection of frequencies with more straightforward identifications; see Appendix \ref{appendix:a} for details.

In Figure \ref{fig:LFcontours} we plot the best-fit indices ($\alpha$) and normalizations at $\nu_u$=600 Hz ($N$) of correlations of the form $\nu_{LF}$=$N(\nu_u/600)^{\alpha}$ fitted in SAX\ J1808, XTE\ J1807 and 4U\ 1820 to the $\nu_{LF}$-$\nu_u$ frequency correlations; these particular fits are from group 2.

For IGRJ 17480, SAX\ J1810 and HETE\ J1900, the contours were not sufficiently constrained in index (in the case of IGR J17480 it was fixed) to plot in Figure \ref{fig:LFcontours}. Instead, we indicate the range in normalization of the $\Delta\chi^2$=9.21 contour with \textit{red, horizontal} lines. 

Similarly, in Figure \ref{fig:Hcontours} we show the $\nu_{h}$-$\nu_u$ correlations fitted in SAX\ J1808, 4U\ 1820 and HETE\ J1900. The fits underlying the contours in Figures \ref{fig:LFcontours} and \ref{fig:Hcontours} are indicated in bold-face in Table \ref{fit}.
In order to compare to the non-pulsating sources, we also plot best-fit indices and normalizations from a selection of sources presented in DK17 in Figures \ref{fig:LFcontours} and \ref{fig:Hcontours} (we omit the ill-constrained source SAX\ J1750). 

We find that for SAX J1808 and XTE\ J1807 the best-fit indices are significantly lower than those of non-pulsating sources 4U\ 1728, 4U\ 1636, 4U\ 1608 and 4U\ 0614 but consistent with the RPM-prediction of 2.0. The normalizations of SAX J1808 and XTE J1807 are much higher than for the non-pulsars, as previously reported by \cite{vanStraaten:2005} based on measurements of characteristic frequencies. 

For HETE J1900 (intermittent pulsar), IGRJ 17480 (11 Hz pulsar) and SAX\ J1810 (non-pulsating) however, the normalizations are consistent with those of the non-pulsars and inconsistent with those of SAX J1808 and XTE J1807. This confirms previous work by \cite{vanStraaten:2005} that not all pulsating sources show shifted centroid frequency correlations with respect to those of non-pulsating sources. We note, however, that this result is based on sparse data, that in the case of IGR J17480 are not concentrated around the $\nu_u$=600 Hz benchmark frequency used to compare the normalizations of frequency correlations in different sources. In that source, a shallower index would also imply a higher normalization. 

The frequency correlations of the non-pulsating source 4U 1820--30 were
previously reported to show a shift similar to (but smaller than) those in the AMXPs SAX\ J1808, XTE\ J0929 and XTE\ J1807 \citep{Altamirano:2005} with respect to the universal correlation for atoll sources known at the time \citep{vanStraaten:2003}. In DK17, using more data, we concluded that the frequency correlations of the sources included in the work by \cite{vanStraaten:2003} significantly differ from each other. Here, we find that the frequency correlations of 4U\ 1820 are similar in index and normalization to those of the non-pulsating sources reported in DK17, as can be seen from Figures \ref{fig:LFcontours} and \ref{fig:Hcontours}.

\begin{figure}
	\includegraphics[width=\columnwidth]{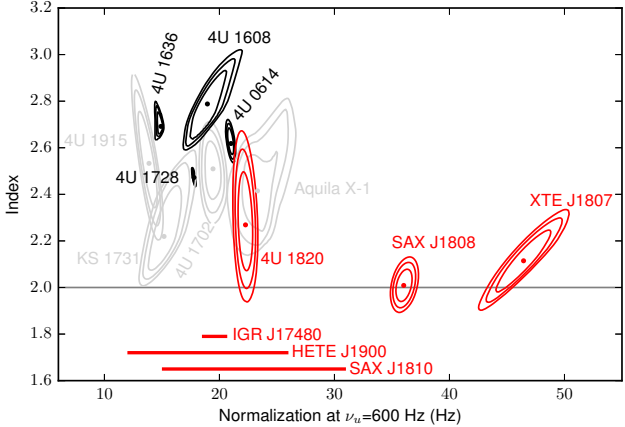}
    \caption{Confidence contours for the best-fit power law index and normalization at $\nu_u$=600 Hz of the $\nu_{LF}$-$\nu_u$ correlations in SAX J1808 (Group 2), XTE J1807 and 4U 1820--30 (Group 2) (this work, \textit{red}) and a selection of non-pulsating atolls (from DK'17, \textit{black, grey}). We plot the 75$\%$ (inner), 95$\%$ (middle) and 99$\%$ (outer) two-parameter confidence limits (corresponding to $\Delta\chi^2$=3.13,6.17 and 9.21, respectively). For IGR J17480, HETE J1900 and SAX J1810, we plot the range in normalization of the $\Delta\chi^2$=9.21 contour in $\textit{red, horizontal}$ lines. The RPM-prediction of index = 2.0 is indicated by the $\textit{dark grey, horizontal}$ line. }
    \label{fig:LFcontours}
\end{figure}

\begin{figure}
	\includegraphics[width=\columnwidth]{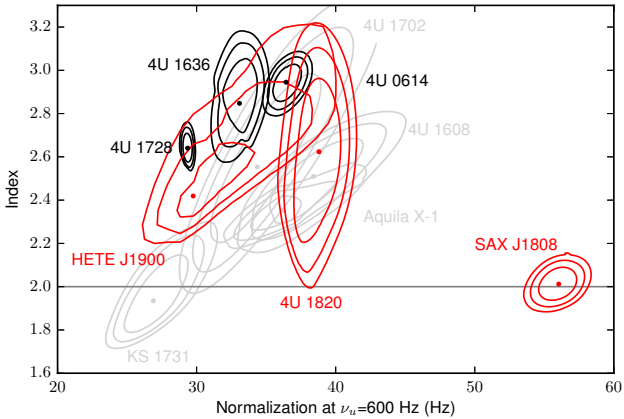}
    \caption{As in Figure \ref{fig:LFcontours}, but for the $\nu_{h}$-$\nu_u$ correlations in SAX J1808, 4U 1820-30 and HETE J1900.}
    \label{fig:Hcontours}
\end{figure}

Our results show that in the ensemble of pulsating and non-pulsating sources, the best-fit indices and normalizations to the frequency correlations for the best-constrained sources systematically and significantly differ 
from each other.

\section{Discussion}
We discuss these results in the context of the RPM, the precessing hot inner flow model and the net orbital precession that results from the combination of magnetic, frame-dragging and classical precession. 

\subsection{Test-particle Lense-Thirring precession}

Although the indices on the best-fit power laws are consistent with the RPM prediction of 2.0, $\nu_{LF}$ and $\nu_h$ in SAX\ J1808 and XTE\ J1807 are much higher than predicted by the model, as discussed in earlier work \citep{Linares:2005, vanStraaten:2005}. Twofold symmetry \citep{Stella:1998} may cause the observed frequency to be twice the LT-precession frequency, and can account for observed frequencies of up to $\sim$24 Hz at $\nu_{spin}$=400 Hz and $\nu_u$=600 Hz, where instead we observe frequencies of $>$37 Hz. It would require unrealistically high values for the NS specific moment of inertia to obtain frequencies this high for the RPM \citep{Stella:1998, Friedman:1986, Cook:1994}.  Moreover, as noted by DK17 the linear spin dependence predicted by the model (if the neutron stars all have similar specific moment of inertia) is not recovered even when adding SAX\ J1808, XTE\ J1807, IGR\ J17480.8, HETE J1900.1 and SAX J1810.8 to the sample of known-spin sources.

\subsection{Precessing torus}
In the hot inner flow model, the torus precession and hence the LF QPO occurs at the disk surface-density weighted average LT-precession frequencies of the orbits within the torus (or twice that frequency), and the kHz QPO is the orbital frequency of the thin disk just outside the torus (at $r_o$) \citep{Ingram:2009, Ingram:2010}.  The higher than test-particle precession frequency is explained by the precession frequency originating closer to the NS than the kHz QPO frequency. The higher index than the test-particle case of 2.0 could be due to narrowing of the torus as the outer disk moves in, as discussed in DK17. As the torus precession frequency is predicted to be less than the LT-frequency close to the inner edge of the torus, clearly in this scenario the highest observed LF QPO frequency can be used to constrain the EOS.

Lower indices in the AMXPs could be due to different inner boundary conditions of the hot flow. For the non-pulsating sources, the inner edge of the torus might be the ISCO or NS surface, both of which are fixed, whereas for the AMXPs this would be expected to be the magnetospheric radius, which moves in with accretion rate at the same time as the outer thin disk moves in. 

\begin{figure*}
	\includegraphics[scale=0.8]{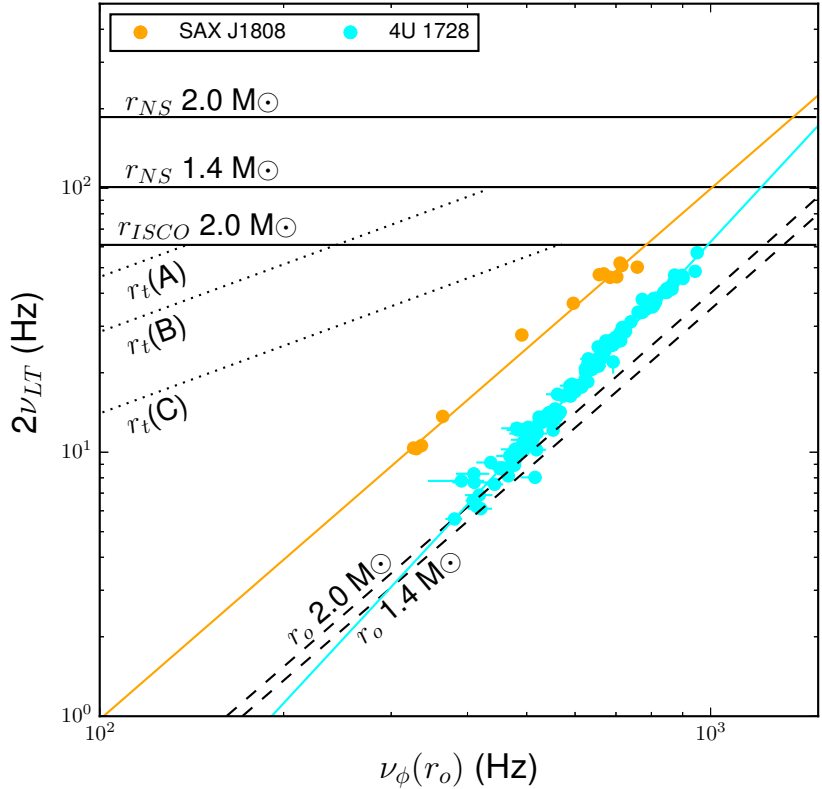}
    \caption{Twice the test-particle LT-precession frequencies (2$\nu_{LT}$) for orbital frequencies ($\nu_{\phi}$) at different radii (magnetic truncation radius $r_t$, \textit{dotted}, stellar and ISCO radius $r_{NS}$ and $r_{ISCO}$, \textit{solid}, inner edge of the accretion disk exterior to the hot inner flow $r_o$, \textit{dashed}) vs. orbital frequency at $r_o$, and $\nu_{LF}$-$\nu_u$ correlations measured by DK17 in for 4U 1728--34 (\textit{cyan}) and SAX\ J1808 (\textit{orange}). To calculate $r_t$ we use parameter sets A, B, and C: ($\frac{M}{M_{\odot}}$, $\frac{R}{\textit{km}}$, $\frac{B_s}{10^8 G}$)=(2.0, 11, 0.5), (1.4, 11.5, 0.5) and (2.0, 11, 1.0). In the precessing torus model $\nu_{LT}$ is predicted to be between the $\nu_{LT}(r_o)$ and min($\nu_{LT}(r_t)$, $\nu_{LT}(r_{NS})$, $\nu_{LT}(r_{ISCO})$) curves for pulsars, and between the $\nu_{LT}(r_o)$ and min($\nu_{LT}(r_{NS})$, $\nu_{LT}(r_{ISCO})$) for non-pulsars.}
    \label{fig:diagram}
\end{figure*}

For illustrative purposes, in Figure \ref{fig:diagram} we plot twice the test-particle LT-precession frequency (2$\nu_{LT}$) at different radii vs. the orbital frequency $\nu_{\phi}$ at the outer edge of the disk (at $r_o$), along with the observed ($\nu_u$-$\nu_{LF}$) frequency correlations of the non-pulsating source 4U 1728--34 ($\nu_{spin}$=363 Hz) reported in DK17 and SAX J1808 ($\nu_{spin}$=401 Hz) (\textit{cyan} and \textit{orange}, respectively). 
We use two representative neutron stars calculated using the APR equation of state (EOS) \citep{Akmal:1998} with $\nu_{spin}$=400 Hz in the RNS code by \cite{Stergioulas:1995} that follows the method by \cite{Cook:1992}: (M=1.4 M$_{\odot}$, R=11.5 km, $j$=0.16) and (M=2.0 M$_{\odot}$, R=11 km, $j$=0.2), where $j$=$Jc/GM^2$ is the spin parameter. We use the Kerr-metric to find the orbital and LT-precession frequencies at various radii\footnote{For the orbital and vertical epicyclic frequencies we use $\nu_{\phi}$=$\nu_K\big($1+$j$($r_g$/$r$)$^{3/2}\big)^{-1}$ and $\nu_{\theta}$=$\nu_{\phi}\big($1-4$j$($r_g$/$r$)$^{3/2}$+3$j$($r_g$/$r$)$^2\big)^{1/2}$, respectively, with gravitational radius $r_g$=$GM/c^2$. These expressions are valid for small disk tilt angles, and  -1$<j<$1. The LT-precession frequency is defined as $\nu_{LT}$=$\nu_{\phi}$-$\nu_{\theta}$ (e.g. \citealt{Klis:2006book}).} We note that many different EOS have been proposed for the central compact star, some of which have been shown to affect the gravitational potential, and hence the orbital frequencies, differently than the APR EOS used in the discussion presented here (i.e. \citealt{Gondek:2014}).

The maximum orbital frequency and hence the maximum $\nu_{LT}$ is limited by the NS surface for the 1.4 M$_{\odot}$ star. For the 2.0 M$_{\odot}$ neutron star, however, it is limited by the ISCO, as r$_{ISCO}$>r$_{NS}$. We indicate the maximum values of 2$\nu_{LT}$ due to the NS surface or ISCO with \textit{black} horizontal line, in Figure \ref{fig:diagram}.

To estimate the truncation radius ($r_t$) of the inner disk due to the magnetic field, we use\footnote{In \cite{Mukherjee:2015} $r_t\approx$(0.01-1)$r_M$ is used to correct for non-dipolar field configurations; numerical simulations for a dipolar field aligned with the spin-axis find $r_t\approx$(0.5-1)$r_M$ \citep{Zanni:2009}. } $r_t\approx$0.5$r_M$, with $r_M$ the Alfv\'en radius:
\begin{align}
r_M=(2G)^{-1/7}B_s^{4/7}R^{12/7}M^{-1/7}\dot{M}^{-2/7}
   \label{eq:Frank1}
\end{align} where $B_s$ is the surface magnetic field strength at the equator, $\dot{M}$ the mass accretion rate and $R$ the stellar radius. Following \cite{Mukherjee:2015}, we use 

\begin{align}
\dot{M}=10^{16}\text{ g s}^{-1} \Bigg(\frac{\epsilon_{\text{bol}}L_X}{1.87\times10^{36} \text{ erg }  \text{s}^{-1}}\Bigg)\times\Bigg(\frac{M}{1.4 \text{ M}_{\odot}}\Bigg)^{-1}\Bigg(\frac{R}{10\text{ km}}\Bigg)
   \label{eq:Frank}
\end{align} with $L_X$=$4\pi d^2F$, and $\epsilon_{bol}$ the bolometric correction factor (which is expected to be $\sim$1-2). Using $L_X\propto\nu_u$ \citep{Wijnands:2001, Bult:2015b}, we find $\nu_{LT}(r_t)\propto\nu_u^{6/7}$. 
In Figure \ref{fig:diagram}, we plot 2$\nu_{LT}(r_t)$ using the observed increase of the X-ray flux (from 40 to 80 mCrab for $\nu_u$ 400-800 Hz) of SAX J1808 over the kHz QPO frequency range, $\epsilon_{bol}$=1.5 and $d$=3.5 kpc \citep{Mukherjee:2015}, for parameter sets A, B, and C: ($\frac{M}{M_{\odot}}$, $\frac{R}{\textit{km}}$, $\frac{B_s}{10^8 G}$)=(2.0, 11, 0.5), (1.4, 11.5, 0.5) and (2.0, 11, 1.0). The magnetic field strength of SAX J1808 is estimated to be $\sim$(0.14--1.8)$\times10^8$ G \citep{Mukherjee:2015}.

The measured frequencies fall within the limits imposed by the maximum LT-precession frequency at the ISCO or stellar surface and magnetosphere, and the minimum LT-precession frequency at the outer edge of the torus, and can therefore be accommodated in the precessing torus scenario. We note that reasonable choices for $M$, $R$, $\epsilon_{bol}$ and $F_X$ show that the truncation radius ($r_t$) for magnetic field strengths above 10$^8$ G becomes too large to explain the observed frequencies in this model. For 4U 1728--34, the weighted average  LT-precession frequency of the torus is close to $\nu_{LT}$ at the outer edge at low $\nu_u$, but at high $\nu_u$ it approaches the LT-precession frequency of orbital motion close to the stellar surface. 

So, in this scenario, the torus must become increasingly centrally concentrated as $\dot{M}$ increases. While this is not expected for a radiation pressure dominated Shakura-Sunyaev disk subject to the LT-torque around BHs \citep{Motta:2018},
interaction of the torus with the NS boundary layer might result in strong changes in the density profile around non-pulsating NSs as a function of $\dot{M}$.
For the AMXPs, interaction with the magnetosphere in the near corotation regime is thought to produce a pile-up of matter just outside the magnetosphere \citep{DAngelo:2010}. The frequency correlation for SAX\ J1808, however, does not suggest any strong changes to the radial density profile of the torus as $\nu_u$ increases.

We conclude that a torus precessing due to frame dragging might explain the observed frequencies in the scenario adopted here if, for the non-pulsating sources, the torus becomes considerably more centrally concentrated with increasing accretion rate. 
We note that this conclusion could in principle also be valid for the model of \cite{Torok:2016}, which identifies the upper and lower kHz QPO in NS-systems with epicyclic modes of an oscillating torus. Analogous to the model of \cite{Ingram:2009}, the torus is expected to precess when it is inclined with respect to the rotation
axis of the central compact object, giving rise to a low frequency oscillation that the LF QPO can be identified with.  

\subsection{Classical and magnetic precession}

In addition to the prograde precession of orbits due to frame dragging ($\nu_{LT}$), the retrograde magnetic and classical precessions are expected to affect orbital motion around a NS (with $\nu_{m}$ and $\nu_{cl}$, respectively). Depending on the dominant effect, net retro- or prograde precession can occur \citep{Shirakawa:2002}. The precession frequencies depend differently on radius and hence on orbital frequency; $\nu_{LT}\propto\nu_{\phi}^2$, $\nu_{m}\propto$ -- ($\nu_{\phi}^{14/3}$), $\nu_{cl}\propto$ -- ($\nu_{\phi}^{7/3}$). 
\\
In DK17, we argued that,  within this framework, indices >2.0 of the power-law correlation of the net precession frequency with orbital frequency can only occur if the magnetic torque is dominant. 
Although both the classical and magnetic precession frequency have a stronger than quadratic dependence on the orbital frequency, the classical precession is expected to be smaller than the LT-precession, leaving only the magnetic precession to produce power law indices >2.0 \citep{Shirakawa:2002, Stella:1998}.

The magnitude of the magnetic torque depends on the misalignment angle ($\theta$) between the magnetic field- and spin axes, the magnetic moment $\mu$ and the mass accretion rate $\dot{M}$. The pulsar-mechanism requires a misalignment between the magnetic field- and spin axes, so that magnetic precession would be expected.\\
It is thought that AMXPs have stronger magnetic fields than non-pulsating sources, as in pulsars the burst oscillations are phase-locked and constant in frequency in the burst tail (e.g., for SAX\ J1808: \citealt{Chakrabarty:2003nat}), whereas the burst oscillations drift in the non-pulsating sources \citep{Muno:2002freq}. 

In \citep{Shirakawa:2002} a global mode analysis of disk precession including magnetic, Lense-Thirring and classical torques is presented. In that work, a stronger magnetic field (2--4$\times10^8$ G)
or a larger misalignment angle ($\sin^2\theta$=0.1--0.5) results in shallower correlations, which matches our findings.  The LT-precession however, dominates the mode frequency in all cases. This leaves the $>2.0$ indices on the frequency correlations in the non-pulsating sources unexplained. 

\section{Conclusions}

We study the correlations of the centroid frequencies of the low frequency (LF) and hump (h) QPOs with that of the kHz QPO in a sample of 6 AMXPs (SAX J1808, XTE J1807, HETE J1900, IGRJ 17480, IGR J17511 and SAX J1748), a burst oscillation source (SAX J1810.8) and a source with unknown spin (4U 1820--30) as observed with RXTE and compare them to those in non-pulsating atoll sources as reported in DK17. For two AMXPs (IGR J17511 and SAX J1748) the relevant QPOs did not occur simultaneously.

The correlations in SAX J1808 and XTE J1807 have significantly lower power law indices than in the non-pulsating sources, and are compatible with the relativistic precession model (RPM) prediction of 2.0. The QPO frequencies are $\sim$4 times higher than predicted by the RPM for NSs with realistic equations of state, however, and we do not find evidence that the best-fit normalizations depend on spin frequency.

We can interpret the frequencies as being due to solid-body like precession of a toroidal hot inner flow, where an average global precession frequency is realized as a surface-density weighted average over multiple radii.
Indeed we find that the observed QPO frequencies are within the range predicted for LT-precession frequencies at the outer radius of the torus as derived from the kHz QPO frequency, $\nu_{LT}$($r_o$), and LT-precession frequencies at the NS-surface, ISCO or magnetosphere, $\nu_{LT}$($r_M$, $r_{NS}$, $r_{ISCO}$). However, to explain the observed frequency correlations, the surface density profile must change as a function of torus size.

Net precession due to combined magnetic, LT- and classical torques could perhaps explain the difference in power law index together with a systematic difference in magnetic field strength between non-pulsating sources and AMXPs. In earlier work by \citep{Shirakawa:2002} on net precession, it was shown that a stronger magnetic field can result in a shallower power law, which matches our result. The steep ($>$2.0) indices on power laws in the non-pulsating sources cannot be explained in this framework, as LT-precession is expected to dominate over the magnetic precession, resulting in a power law index $<$2.0.

The data on the other three sources (the two AMXPs HETE J1900 and IGRJ 17480, and SAX J1810.8) do not allow any strong constraints on the frequency correlation indices and normalizations. They seem to be inconsistent with those of SAX J1808 and XTE J1807 and instead similar to those of the non-pulsating sources, but more data is needed to draw strong conclusions. 4U\ 1820--30 shows frequency correlations that are compatible with those of other non-pulsating sources. We conclude that the frequency correlations in NS-LMXBs with kHz and LF-QPOs vary significantly, and that this variation can be interpreted in the context of models involving (rigid toroidal) precession of orbits in the vicinity of the NS due to Lense-Thirring, classical, and magnetic precession, extending to NSs recent findings for LF-QPOs in BH systems \citep{Ingram:2016}. It would be interesting to investigate using GRMHD simulations whether stable torus precession could occur due to the combined torques expected around a neutron star, and how the surface density profile and geometry of the inner accretion flow might evolve as the source state changes.

\section*{Acknowledgements}
This research has made use of data obtained through the High Energy Astrophysics Science Archive Research Center Online Service, provided by the NASA/Goddard Space Flight Center.
This work is (partly) financed by the Netherlands Organisation for Scientific Research (NWO). 
We thank the referee for the constructive feedback and suggestions that improved the quality of the paper.




\bibliographystyle{mnras}

\bibliography{Bibliography} 




\appendix
\label{appendix:a}
\section{Results for individual sources}
Here we present detailed results for the individual sources included in our sample. Contrary to some previous works (e.g. \citealt{Bult:2015b, vanStraaten:2005}) we use centroid frequency $\nu_{0}$ instead of characteristic frequency $\nu_{max}$ as a fit parameter. Broad, zero-centered Lorentzians are therefore absent from our frequency-frequency plots. 

\subsection{SAX J1808.4--3658}
For SAXJ1808 we rely on the data selection presented in \cite{Bult:2015b} hereinafter BK15, and use the data interval numbering introduced in that paper. Analysis choices different from that work are outlined below. 
We use energy channels covering the 2-20 keV band. 

\begin{figure}
	\includegraphics[width=\columnwidth]{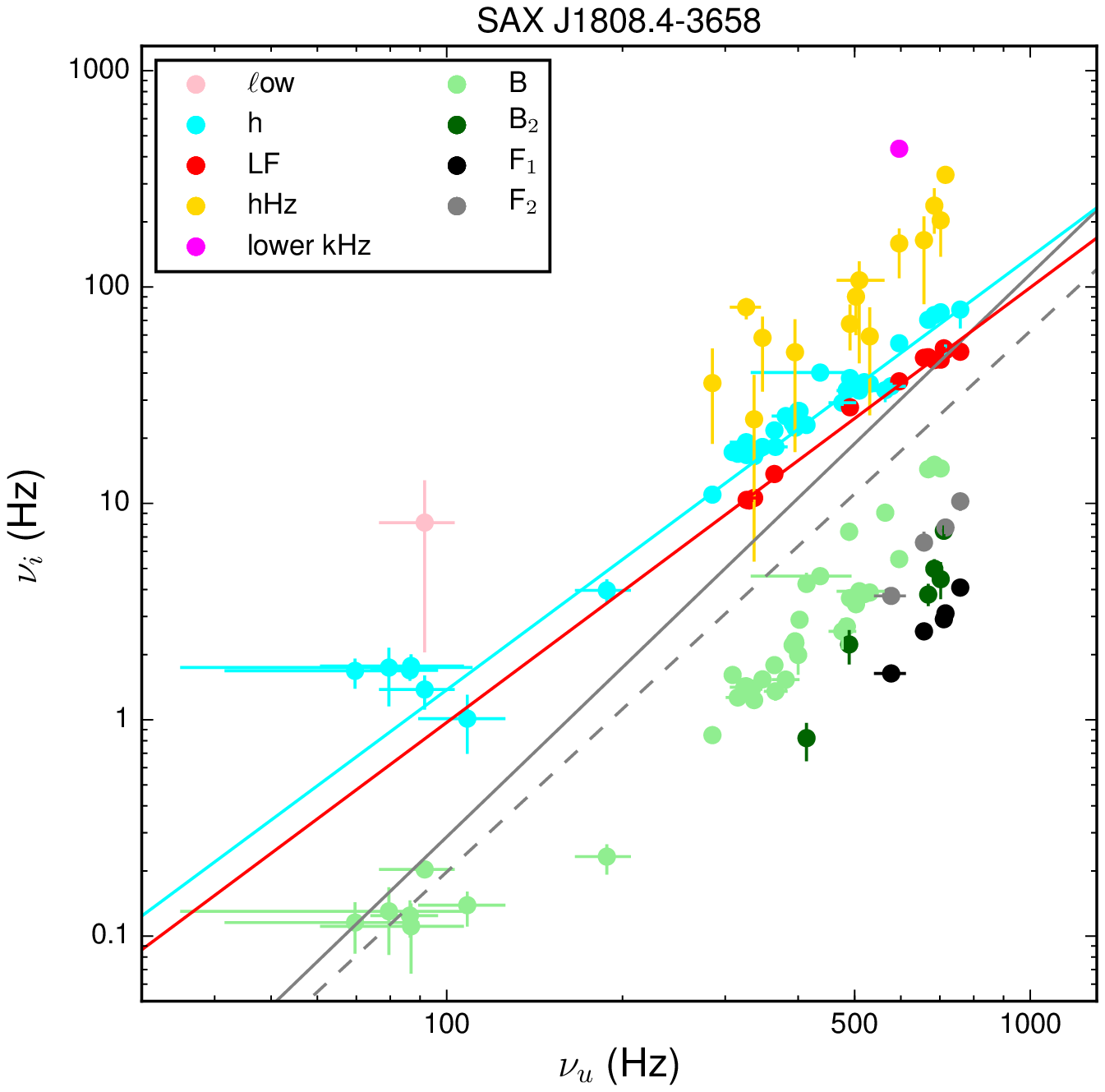}
    \caption{Centroid frequencies of power spectral components  fitted in SAX J1808 vs. $\nu_u$. We plot the best-fit power laws to $\nu_{LF}$ vs. $\nu_u$ (\textit{red line}) and $\nu_h$ vs. $\nu_u$ (\textit{cyan line}). We also plot $\nu_{LF}$ vs. $\nu_u$ (\textit{grey, dashed line}) and $\nu_h$ vs. $\nu_u$ (\textit{grey, solid line}) for non-pulsating atoll source 4U\ 1728--34 (DK17) for comparison. For clarity, we omit centroid frequencies with large error bars.}
    \label{fig:freqSAX}
\end{figure}

In the power spectra of intervals 6, 10, 11, 12, 13 (all from 1998), 1 (2005), 1,6,7 (from 2008), and 2,7,8,9, 10 (2011), the upper kHz QPO centroid frequency has large errors, is fixed at zero, or the QPO is not detected. We omit frequencies from these intervals from our analysis. 
L$_{\ell ow}$ is a zero-centered Lorentzian in all but one fit, explaining the absence of this feature in Figure \ref{fig:freqSAX}. 

We detect the LF QPO ($>$3$\sigma$, $\nu_0\sim$10 Hz) in intervals 6, 7, 8 and 13 of the 2002 outburst, while in BK15 this QPO is not reported. (We note that the frequency of this QPO, 10.6$\pm$0.1 Hz in interval 6, is close to, but higher than the frequency needed ($\sim$8 Hz) to produce the simultaneously present "410 Hz QPO" -- at 409.6$\pm$0.6 in the averaged power spectrum of intervals 6,7 and 8; BK'15 -- in a retrograde beat with the 401 Hz spin frequency. We do not investigate the 410 Hz QPO further in this work.)

\begin{figure}
	\includegraphics[width=\columnwidth]{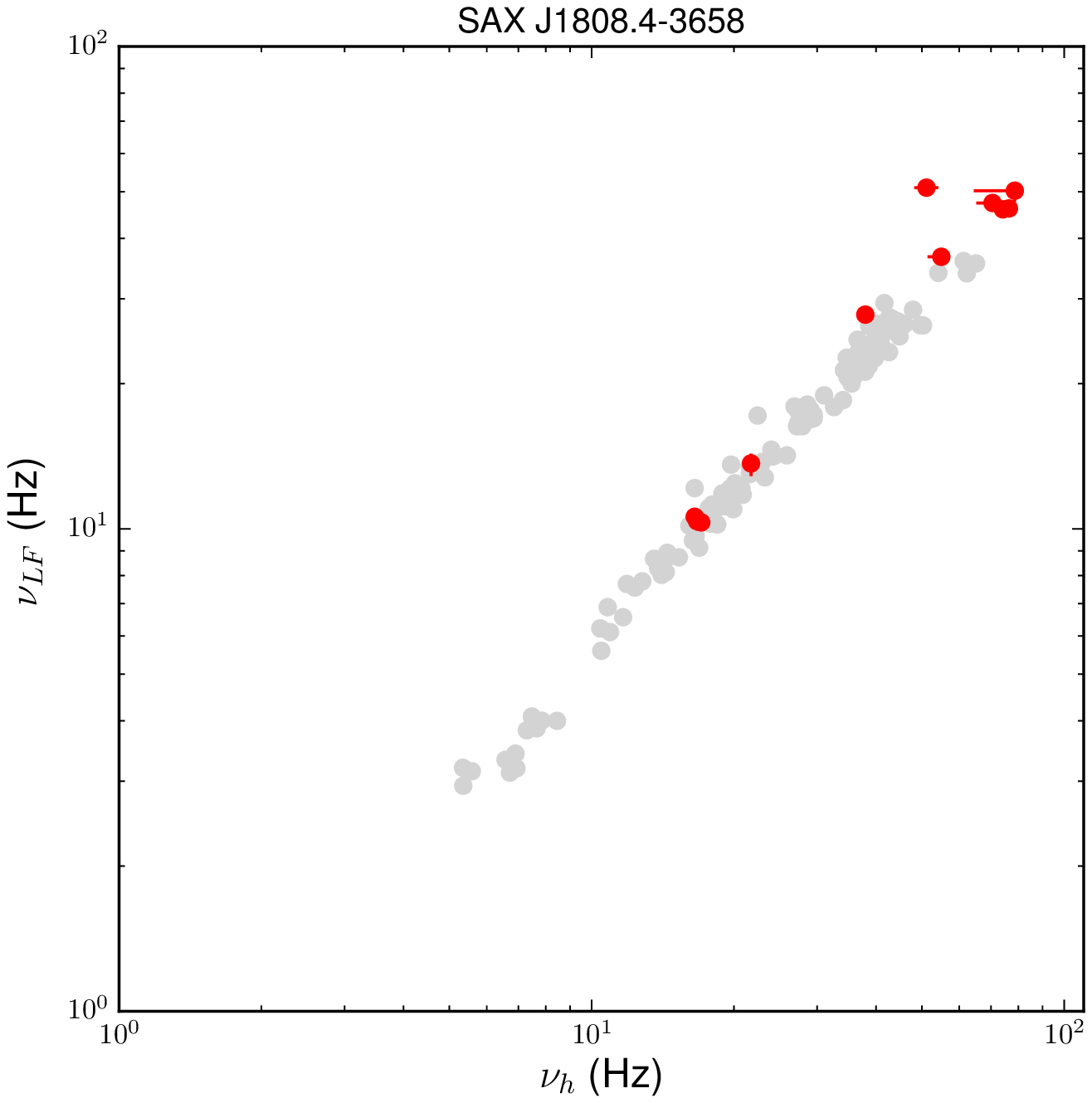}
    \caption{Centroid frequencies; $\nu_{LF}$ vs. $\nu_{H}$ fitted in SAX J1808. For comparison, we plot $\nu_{LF}$ vs. $\nu_{h}$ fitted in power spectra of all non-pulsating atoll sources from DK17 in \textit{grey}.}
    \label{fig:LFHSAX}
\end{figure}

Like BK15, we identify the Lorentzian with $\nu_0\sim$25-40 Hz in power spectra with $\nu_u$ between $\sim$500-600 Hz as L$_h$. It is possible, however, that the Lorentzian is a blend of L$_{LF}$ and L$_h$, such as seen for similar $\nu_u$ values in non-pulsating atoll sources (DK17). As $\nu_{LF}<\nu_h$, such blending could explain the slightly lower $\nu_h$ values compared to the non-pulsating sources (\ref{fig:LFHSAX}).

\begin{figure}
	\includegraphics[width=\columnwidth]{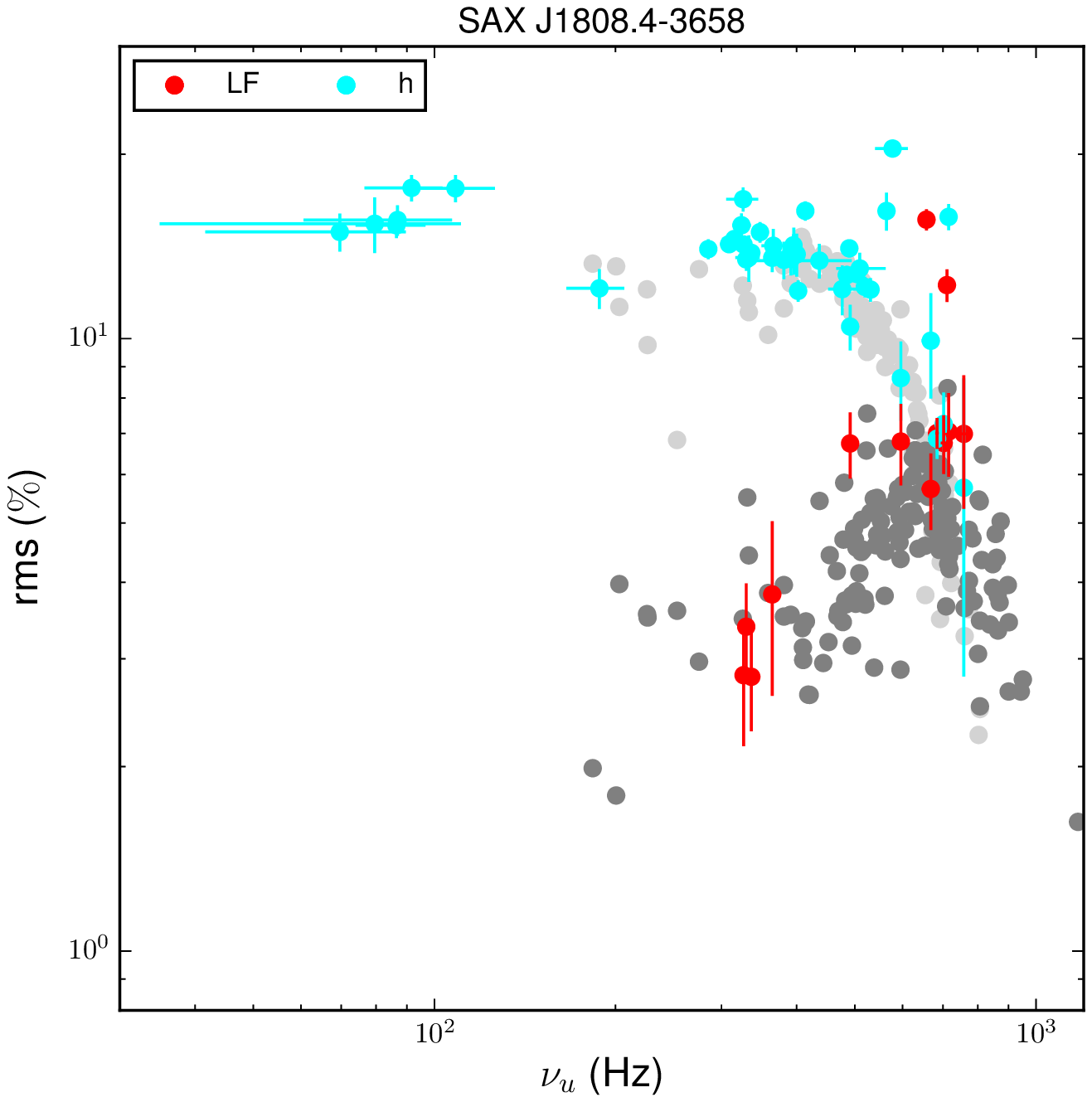}
    \caption{Rms$_{LF}$ and rms$_{h}$ vs. $\nu_u$ fitted in SAX J1808. For comparison, we plot rms$_{LF}$ and rms$_{h}$ vs. $\nu_u$ fitted in power spectra of 4U 1728--34 from DK17 in $\textit{dark}$ and $\textit{light grey}$, respectively.}
    \label{fig:IPSAXu}
\end{figure}

\begin{figure}
	\includegraphics[width=\columnwidth]{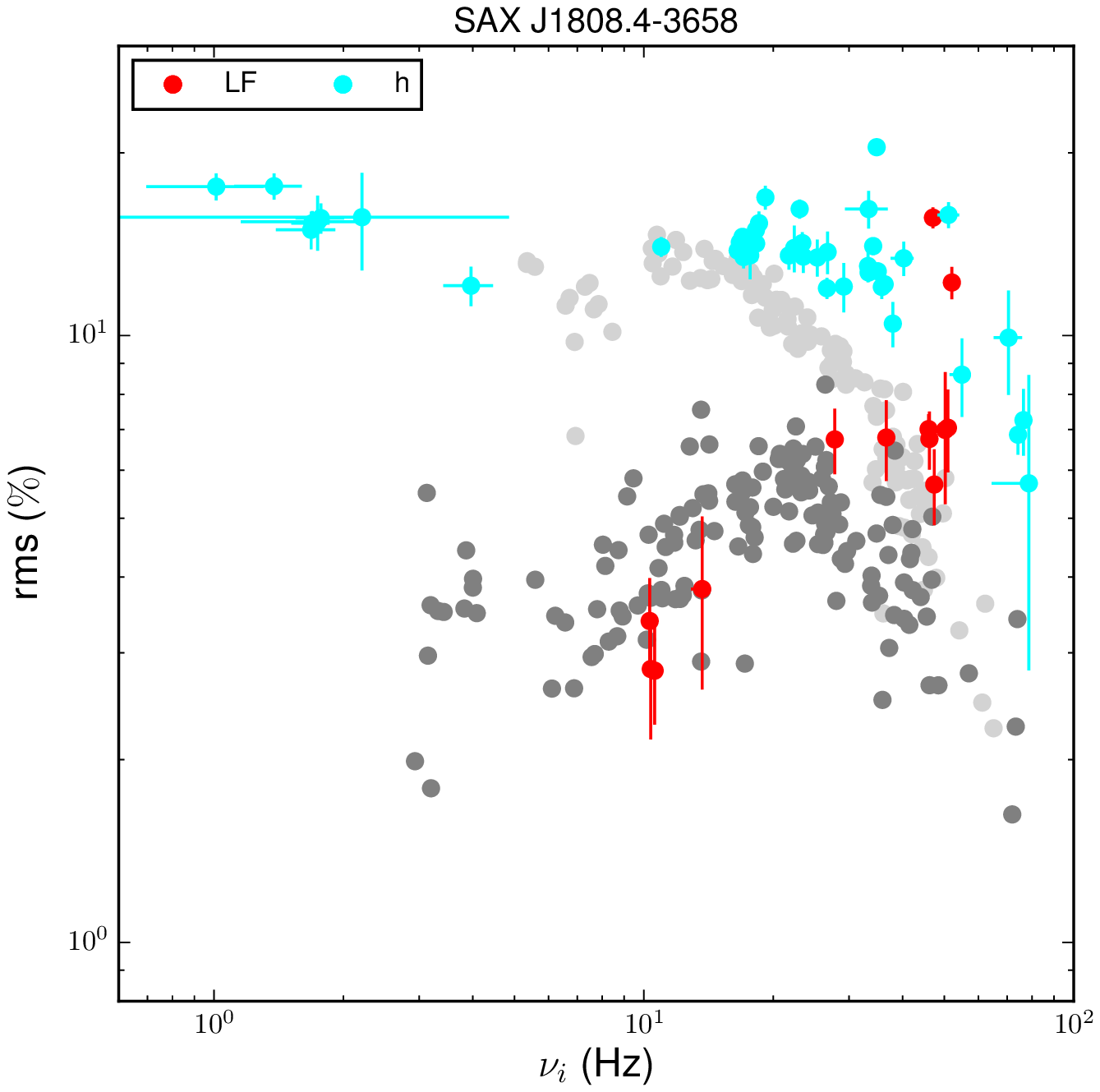}
    \caption{Rms$_{LF}$ vs. $\nu_{LF}$ and rms$_{h}$ vs. $\nu_h$ fitted in SAX J1808. For comparison, we plot rms$_{LF}$ vs. $\nu_{LF}$ and rms$_{h}$ vs. $\nu_h$ fitted in power spectra of 4U 1728--34 from DK17 in $\textit{dark}$ and $\textit{light grey}$, respectively.}
    \label{fig:IPSAX}
\end{figure}

In intervals 1 and 2 of the 2011 outburst ($\nu_u\sim$600 Hz), we identify the Lorentzian with $\nu_0\sim$40-50 Hz as L$_{LF}$, contrary to BK15 who identify it as L$_h$. In the very similar power spectrum of interval 3 (also from 2011), we and BK15 both fit two Lorentzians to this feature, identifying the more dominant and narrower lower frequency peak as L$_{LF}$, and the other feature as L$_h$. The single Lorentzian fitted to the power spectra in intervals 1 and 2 presumably mostly represents this lower, dominant peak. In Figures \ref{fig:IPSAXu} and \ref{fig:IPSAX} the two high values of rms$_{LF}$ originate from this blended power spectral component. BK15 note that L$_h$ and L$_{LF}$ might be affected by the low-luminosity flaring feature present below $\sim$10 Hz. When we plot $\nu_{LF}$ vs. $\nu_h$ (Figure \ref{fig:LFHSAX}), this particular power spectral fit produces an outlier relative to the $\nu_{LF}$-$\nu_h$ correlation seen for other intervals of SAX\ J1808 and for the non-pulsating sources, supporting this view.

The hump feature reported in BK15 for interval 4 of the 2011 outburst was not required in our fit; we therefore omit this feature. We identify the single Lorentzian as L$_{LF}$, as in intervals 1 and 2.  
Likewise, a lower kHz QPO was not required in intervals 4 (2002) and 5 (2011); instead we fit a slightly broader hHz component.
As mentioned earlier, we use multiple Lorentzians to fit the asymmetric flaring feature in intervals 5 (2008) and 1,3,4 (2011), which we name  L$_{F_n}$ in Table \ref{tab:pars2} and Figure \ref{fig:freqSAX}. 

Clearly then, in SAX J1808 identification of features is not straightforward.  Blending appears to play a significant role. 
In several power spectra, however, both L$_h$ and L$_{LF}$ are simultaneously detected. 
Their frequencies correlate with each other as in non-pulsating atoll sources, see Figure \ref{fig:LFHSAX}, and the strength, width and frequency of the QPOs in the power spectra match that of non-pulsating sources, albeit at lower $\nu_u$. We presented the best-fit power law parameters to two different data groups in Table \ref{fit}, one comprising all centroid frequencies of L$_h$ and L$_{LF}$ identified as described above (Group 1), the other comprising only the centroid frequencies of L$_h$ and L$_{LF}$ fitted simultaneously (Group 2, these power laws are plotted in Figure \ref{fig:freqSAX} in \textit{cyan} and \textit{red}, respectively) where blending is less of an issue. Group 2 is comprised of intervals 2, 3, 4, 5, 6, 7, 8, 13 (all from 2002), 7 (2005) and 5 (2011). We omit interval 3 (2011), because of the tentative identification of L$_{LF}$ and L$_h$ described above. 

We note that for SAX\ J1808, plotting rms$_{LF,h}$ vs. $\nu_{u}$ (Figure \ref{fig:IPSAXu}) matches what is seen in non-pulsating sources (plotted in $\textit{grey}$) better than plotting rms$_{LF,h}$ vs. $\nu_{LF,h}$ (Figure \ref{fig:IPSAX}). While the $\nu_{LF,h}$-$\nu_u$ correlations for SAX J1808 do not overlap with those of the non-pulsating sources, the rms$_{LF,h}$-$\nu_u$ plots do. This suggests that the LF,h QPO strength and frequency shares a common physical origin with kHz QPO frequency, but that the LF,h QPO frequency is also affected by another process.

\subsection{XTE J1807-294}

The timing behaviour of XTE J1807--294 is described in \cite{Linares:2005}. We use the data groups defined in that work,
but apply a 2-20 keV energy selection after verifying that this selection improves the detection significance of power spectral features. 

\begin{figure}
	\includegraphics[width=\columnwidth]{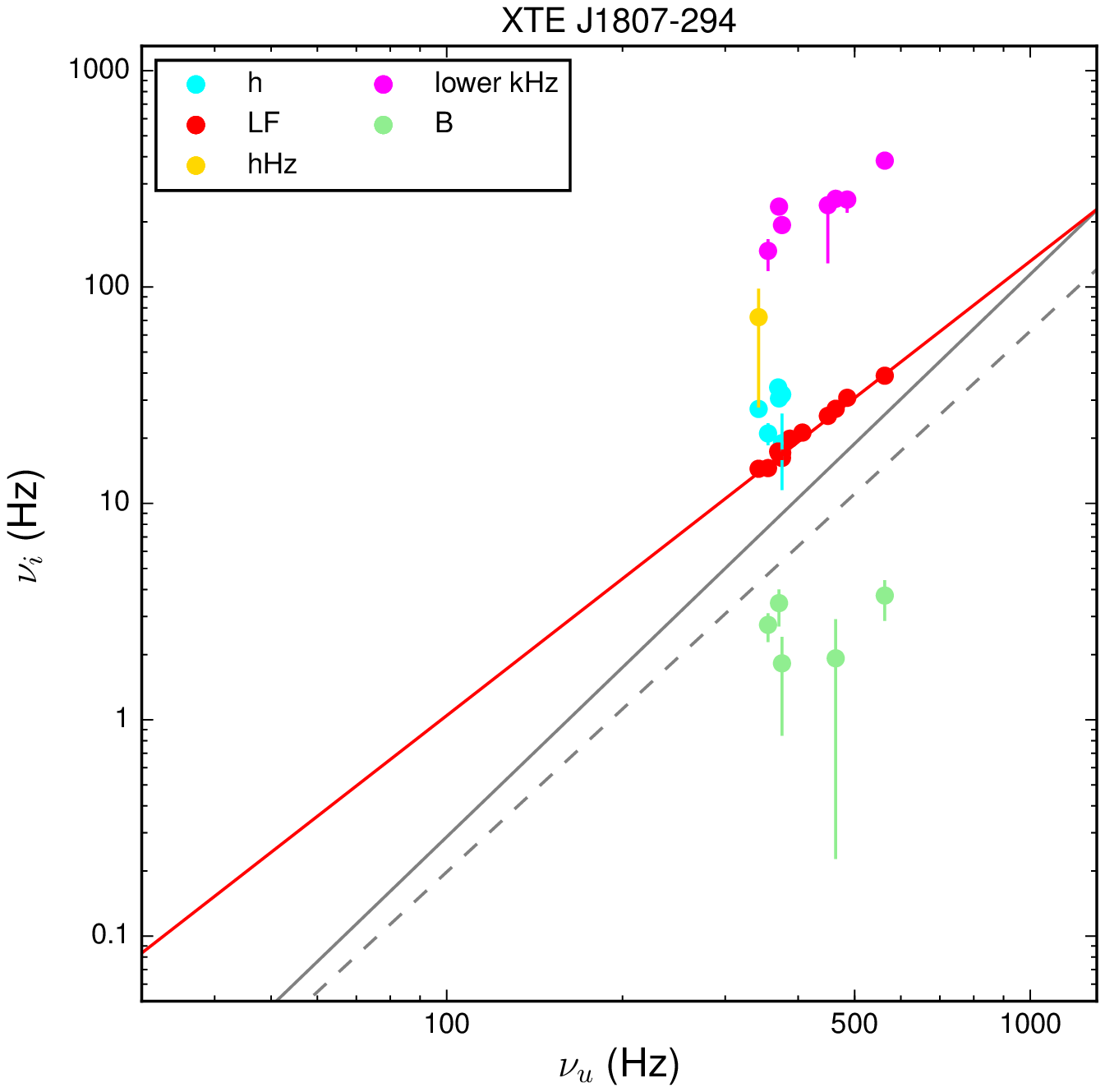}
    \caption{As in Figure \ref{fig:freqSAX}, but for XTE J1807. }
    \label{fig:freqXTE}
\end{figure}

\begin{figure}
	\includegraphics[width=\columnwidth]{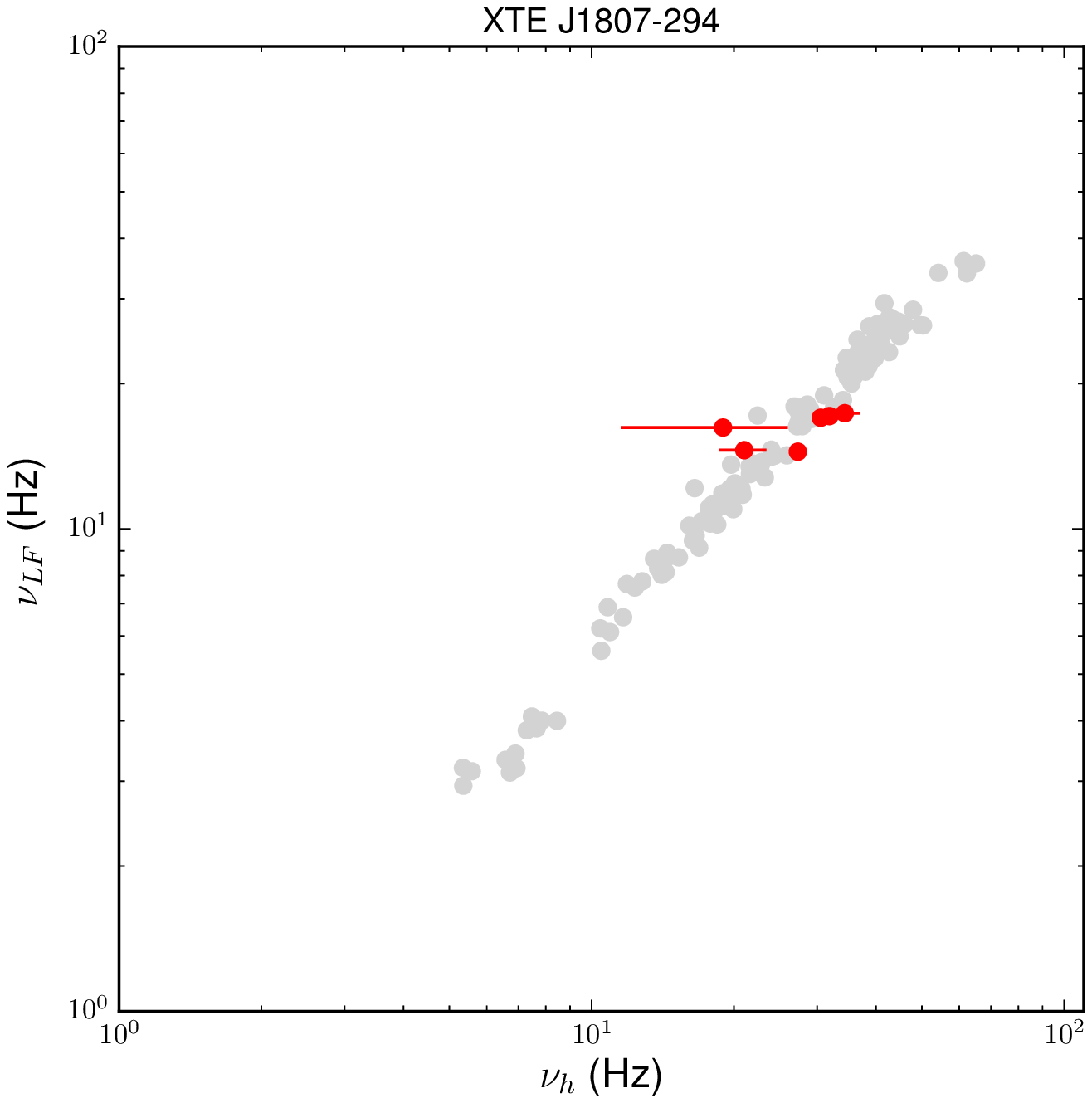}
    \caption{As in Figure \ref{fig:LFHSAX}, but for XTE J1807. }
    \label{fig:LFHXTE}
\end{figure}

We fit the average power spectra of all observations in groups A, E, F, G and H. In group B we fit 80145-01-03-03 separately, as L$_h$ is a more prominent feature in this observation than in others. We fit the power spectra of all observations in groups C and D separately. The identification of features is the same as described in \cite{Linares:2005}.
 In Figure \ref{fig:freqXTE} we plot $\nu_{LF}$ and $\nu_h$ vs. $\nu_u$, and in Figure \ref{fig:LFHXTE} we plot $\nu_{LF}$ vs. $\nu_{h}$. 

As L$_h$ is either absent or zero-centered in most power spectra, we only present the power law fit to the $\nu_{LF}$-$\nu_u$ correlation.   

\begin{figure}
	\includegraphics[width=\columnwidth]{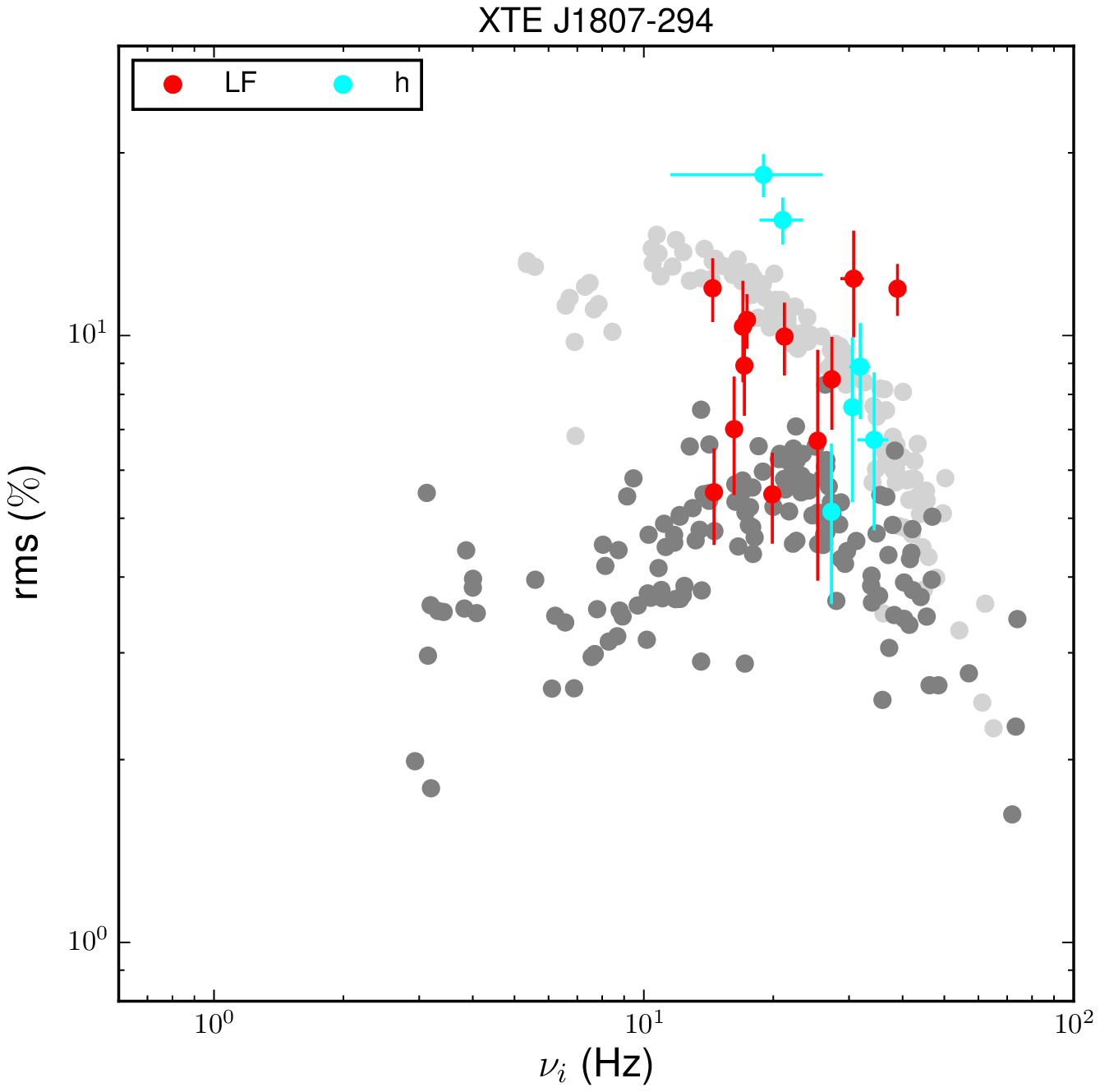}
\caption{As in Figure \ref{fig:IPSAX}, but for XTE J1807. }
    \label{fig:IPXTE}
\end{figure}

\subsection{HETE J1900.1--2455}

\begin{figure}
	\includegraphics[width=\columnwidth]{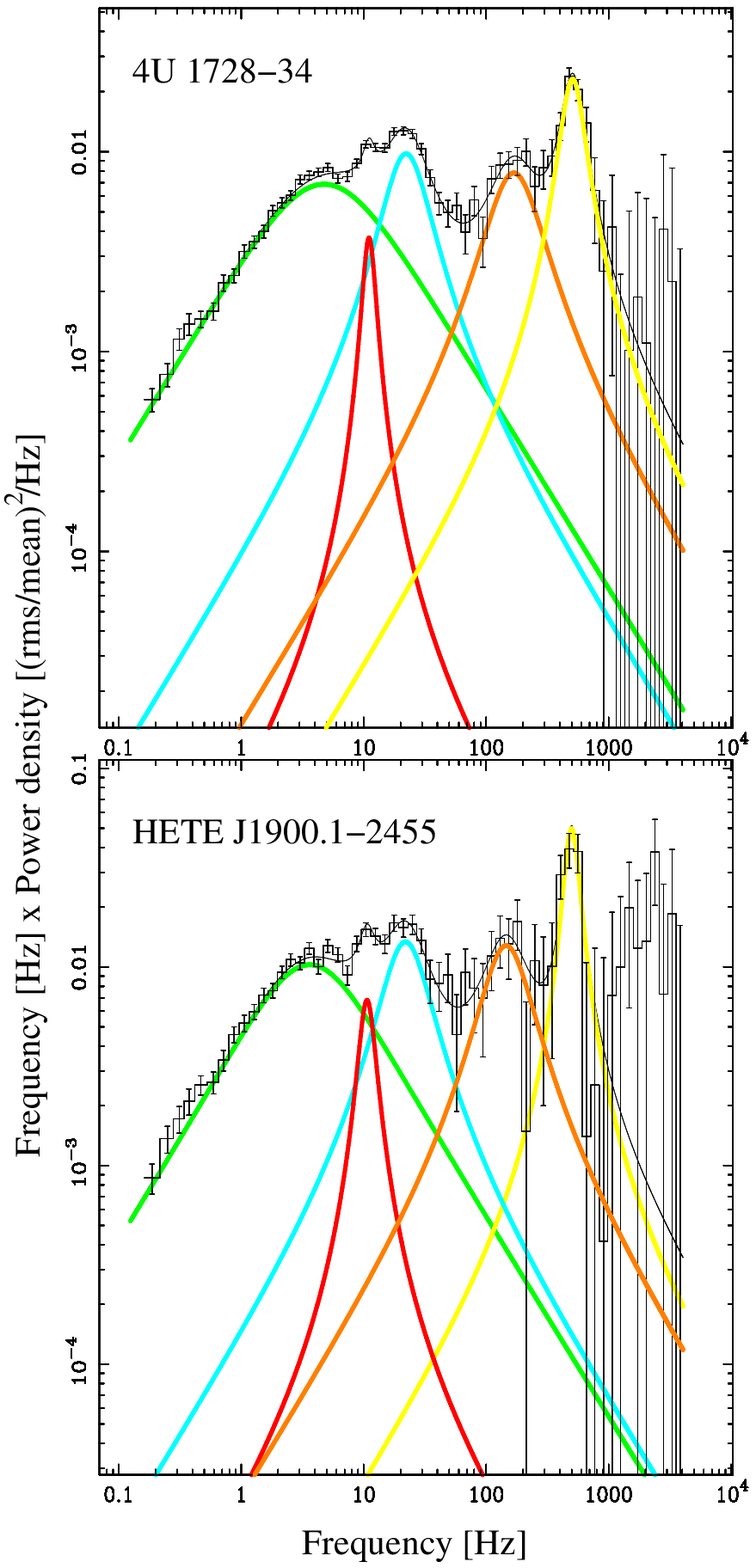}
    \caption{Power spectra of 4U\ 1728--34 (top, ObsID 50023-01-25-00, from DK17) and HETE\ J1900.1 (bottom, ObsID 91015-01-04-06). 
    Pulsations at 377.3 Hz are present in this observation of HETE J1900.1; the pulsar spike was removed. For 4U 1728--34 and HETE J1900.1, $\nu_u$ is 494$\pm$12 (7.8$\sigma$) and 493$\pm$16 Hz (4.8$\sigma$), $\nu_{LF}$ ($\textit{red}$) is 11.0$\pm$2.7 Hz (3.6 $\sigma$) and 10.5$\pm$4.5 Hz (2.4 $\sigma$), $\nu_h$ ($\textit{cyan}$) is 20.0$\pm$7.4 (12$\sigma$) and 19.5$\pm$1.9 Hz (4.8 $\sigma$), respectively.}
    \label{fig:HETE-1728}
\end{figure}
Following \cite{Patruno:2017} we use energy channels 5-37, and calculate 128-second FFTs. 
Our results are slightly different from that earlier work due to analysis choices described below. 

While the significance of broad features generally improves when averaging observations; the LF-QPO cannot be resolved in the averaged power spectra used by \cite{Patruno:2017}, who instead fit a single broad Lorentzian they identify as L$_{h}$. Although marginally, we detect L$_{LF}$ and L$_h$ simultaneously in several single observations, see the bottom panel of Figure \ref{fig:HETE-1728} for an example. We therefore subdivided some of these groups. 

For group 2, we find that two different non-zero centered Lorentzians (L$_{LF}$ and L$_{h}$) can be fitted in two (91015-01-04-04 and 91015-01-04-06) of the three individual observations. The third (91015-01-04-07) is different from the other two (no kHz QPO, hump at much lower frequency, higher hard color, lower soft color by about 5 per cent, and observed 3 days earlier than the next observation (91015-01-04-04) in the group).\\
For these reasons, we use the frequencies fitted in 91015-01-04-04 and 91015-01-04-06, and omit 91015-01-04-07 from our analysis.

In group 3, which contains a single observation, we fit both L$_{LF}$ ($\nu_0$=17 Hz) and L$_{h}$ ($\nu_0$=27 Hz), where \cite{Patruno:2017} report a single blended feature, L$_h$ ($\nu_0$=22 Hz). 

L$_h$ is broadened in the average power spectrum of the four observations in group 4. We choose to combine the similar power spectra of 91015-01-07-00 and 91015-01-06-01, which results in fitting a narrower L$_h$.
We fit  91015-01-06-02 separately (L$_h$ is zero centered), and we omit 91015-01-07-01, due to the absence of the kHz QPO in this observation. \\
In group 5, we combine 91059-03-03-03 and 91059-03-03-02. The upper kHz QPO is detected in this data selection at $>$3$\sigma$, whereas in the average of all six observations in group 5, (6 obs), we detect it at only 1.1$\sigma$. We omit the rest of the observations in this group as no kHz QPO is detected.

Finally, we adopted groups 1, 6, 7, 8, 9, 10, 11, 12 without change. Very broad kHz QPOs were fitted in Groups 1,7, 10 and 11, so that $\nu_u$ has large errors; we omit these from our correlation fitting. 

\begin{figure}
	\includegraphics[width=\columnwidth]{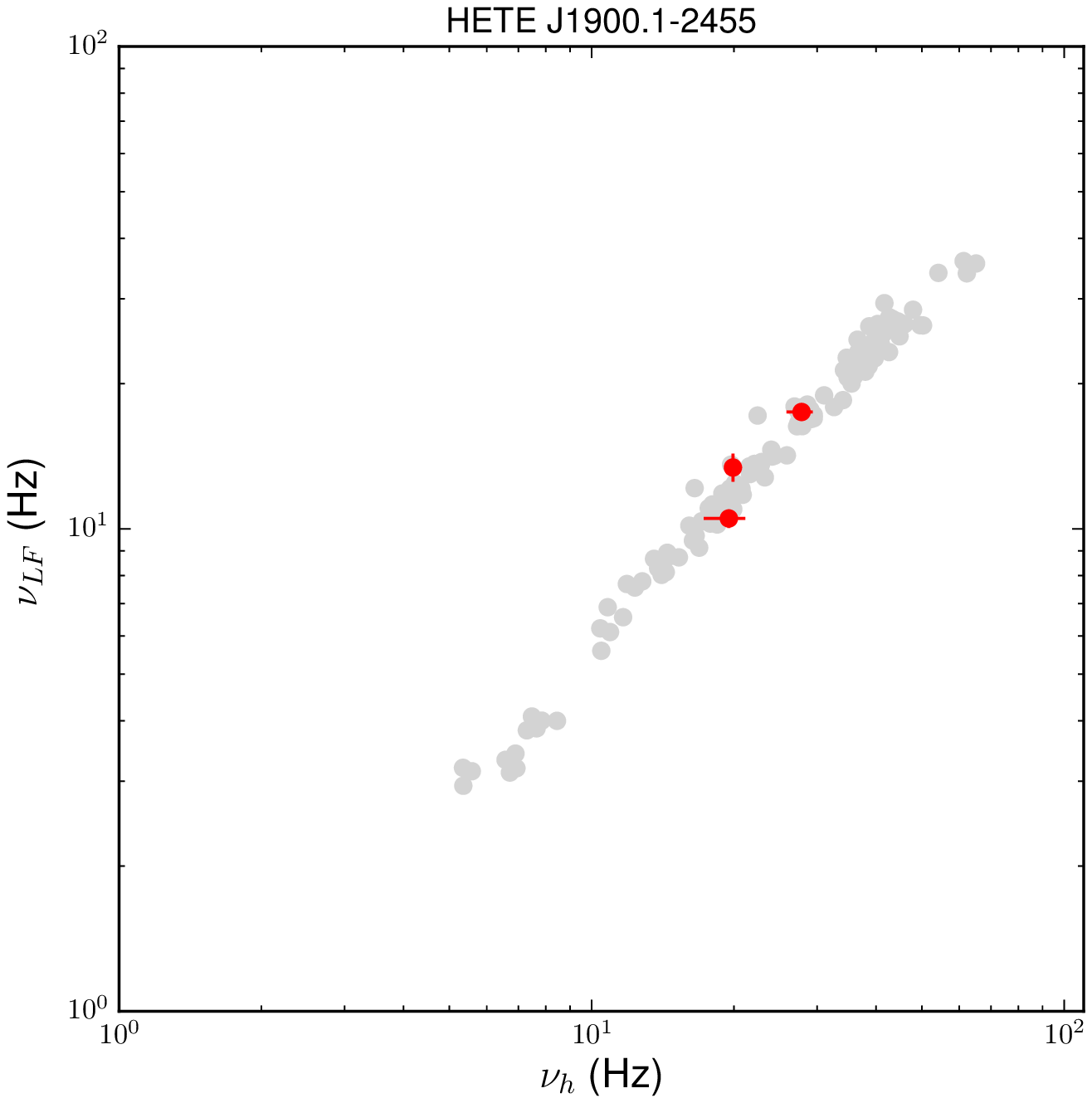}
    \caption{As in Figure \ref{fig:LFHSAX}, but for HETE J1900. }
    \label{fig:LFHHETE}
\end{figure}

Although the simultaneously detected L$_{LF}$ and L${_h}$ are $<$3$\sigma$ ($\sim$2.5$\sigma$), when plotting $\nu_{LF}$ vs. $\nu_{h}$ (see Figure \ref{fig:LFHHETE}) the points fall on the correlation seen in the non-pulsating atolls. 
The frequency correlations involving $\nu_u$ in HETE J1900 are not smooth. The data quality does not allow for any strong conclusions on the frequency correlations, but we do provide constraints on the power law normalization and index in Table \ref{fit}.

\begin{figure}
	\includegraphics[width=\columnwidth]{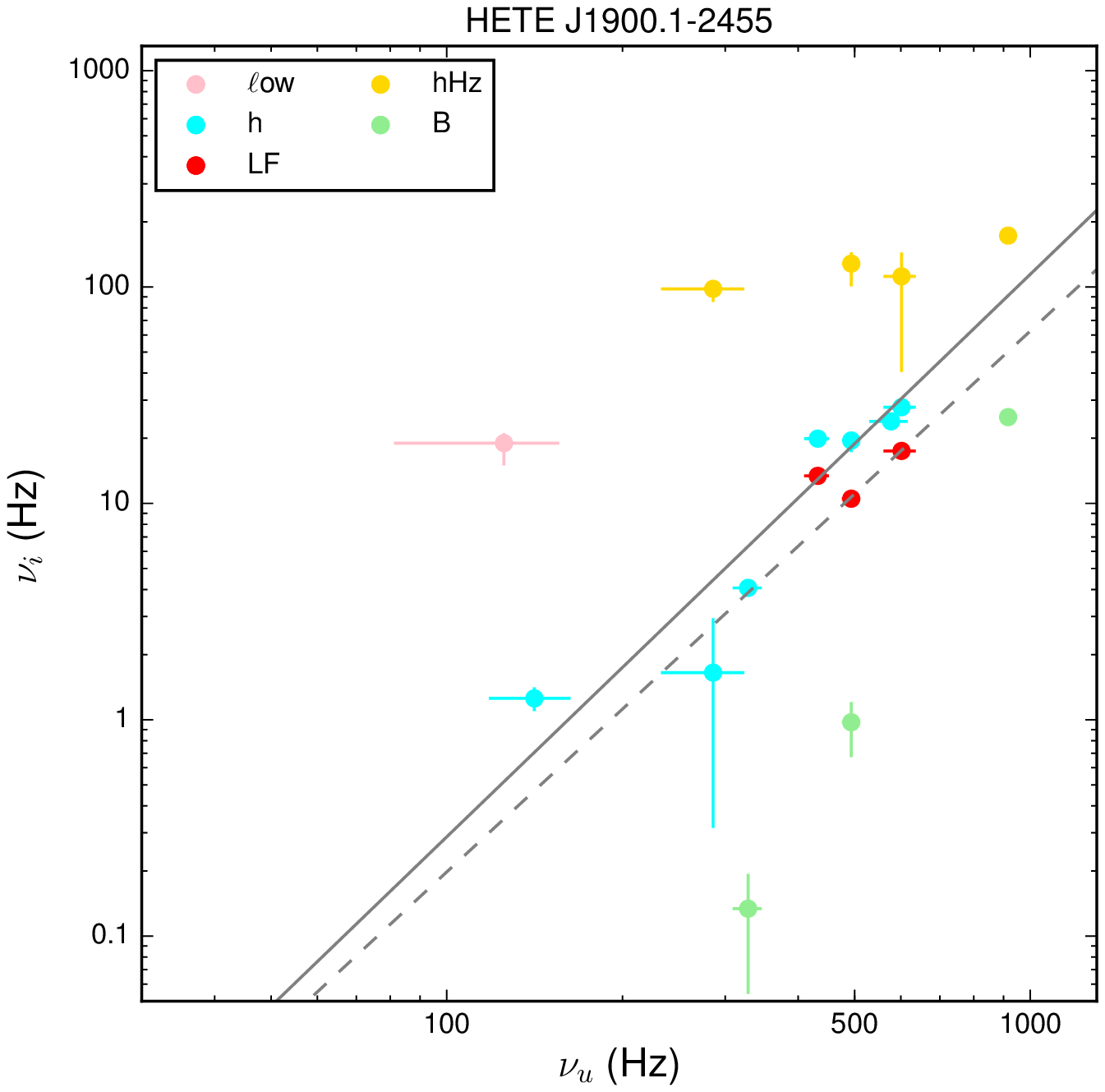}
    \caption{As in Figure \ref{fig:freqSAX}, but for HETE J1900.  }
    \label{fig:freqHETE}
\end{figure}

\begin{figure}
	\includegraphics[width=\columnwidth]{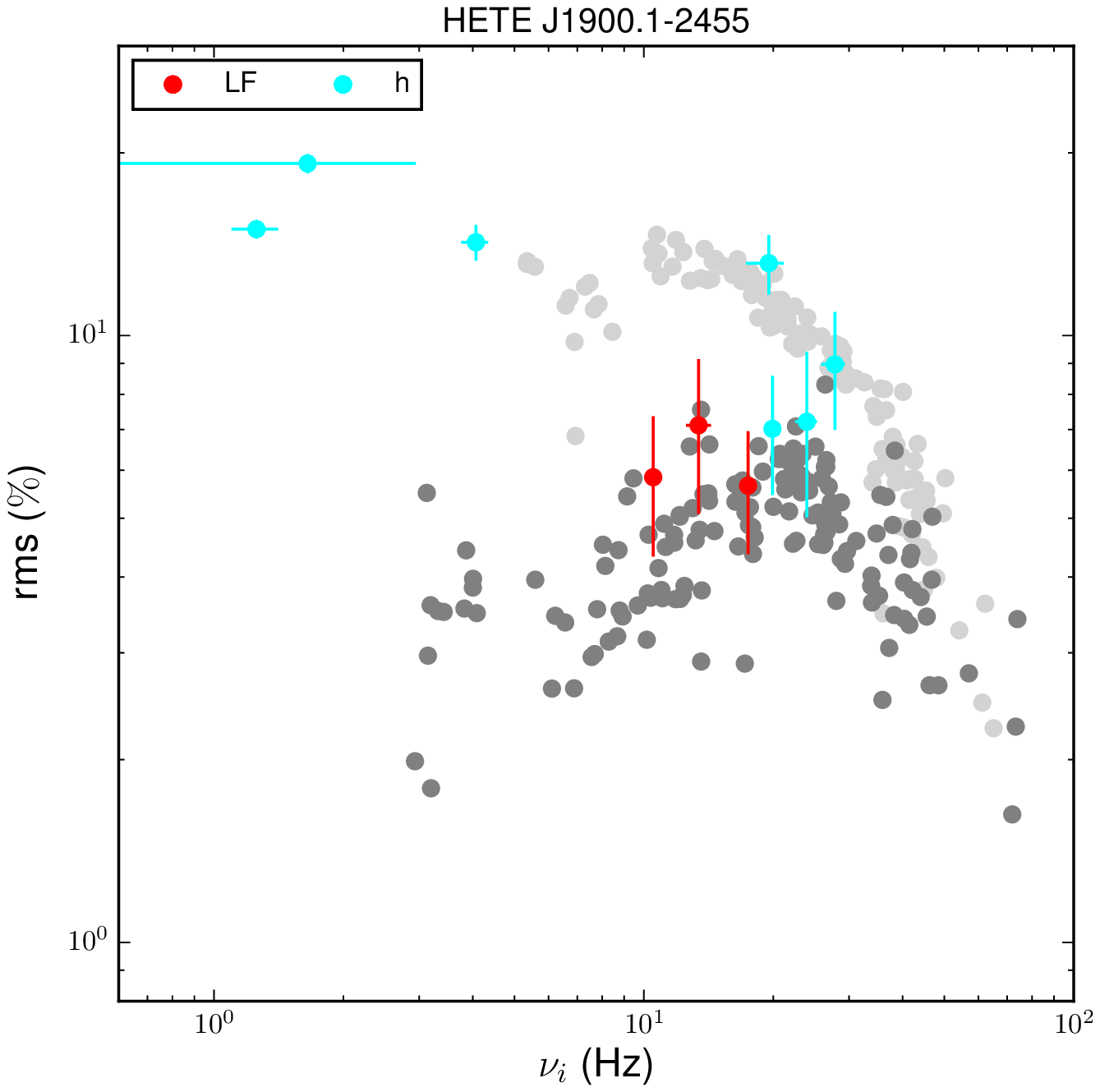}
\caption{As in Figure \ref{fig:IPSAX}, but for HETE J1900. }
    \label{fig:IPHETE}
\end{figure}

\subsection{IGR J17480--2446}
\begin{figure}
	\includegraphics[width=\columnwidth]{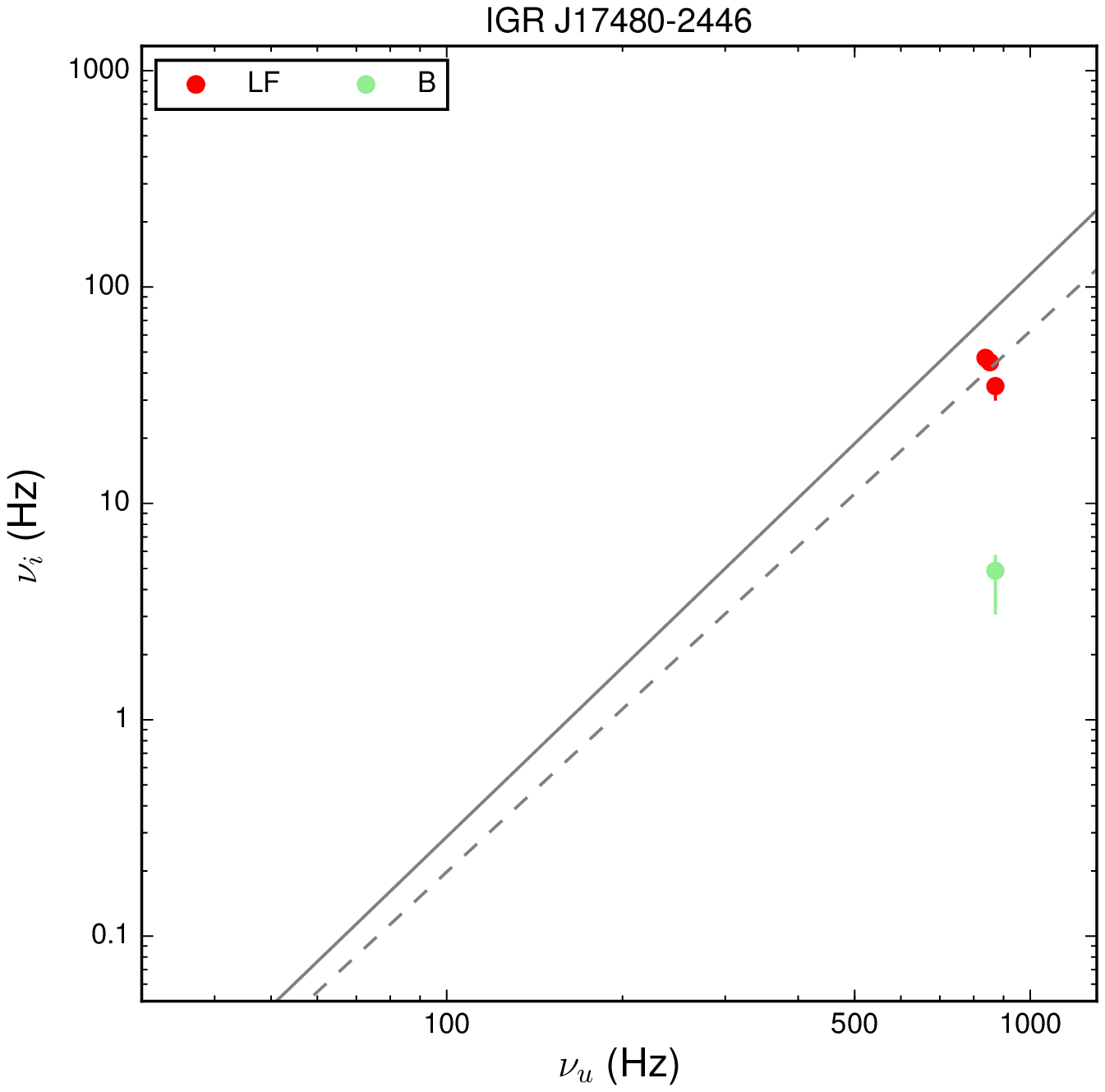}
    \caption{As in Figure \ref{fig:freqSAX}, but for IGR J17480--2446. }
    \label{fig:freqter5}
\end{figure}

\begin{figure}
	\includegraphics[width=\columnwidth]{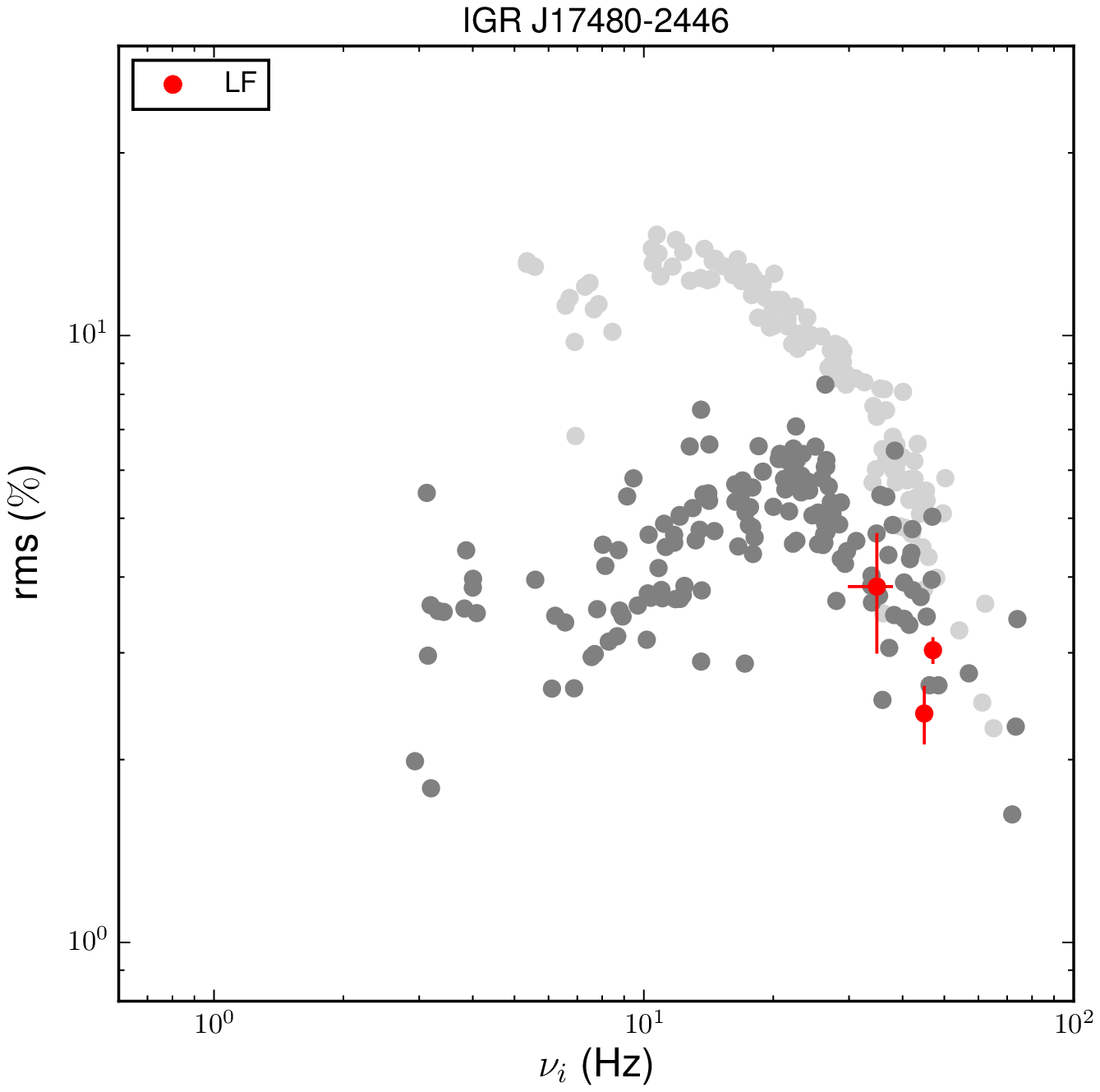}
    \caption{As in Figure \ref{fig:IPSAX}, but for IGR J17480--2446.}
    \label{fig:IPter5}
\end{figure}

IGR J17480--2446, or Terzan 5-X2, is an 11 Hz pulsar.
We use the same data set as 
\cite{Altamirano:2012}, and use all energy channels available as the signal-to-noise is low. In Figure \ref{fig:freqter5} we plot $\nu_{LF}$ vs. $\nu_u$; the data are consistent with the correlations of 4U\ 1728--34. The rms$_{LF}$ vs. $\nu_u$ plot is shown in Figure \ref{fig:IPter5}. Due to the scarcity and concentration of measurements of $\nu_{LF}$ at $\nu_u\sim$900 Hz, we fix the index of the power law we fit to 2.4 to constrain the normalization.
 Fixing it to 2.0, as in SAX J1808, resulted in a similar normalization of $N=$22.5$\pm$0.4 (with similar reduced $\chi^2$).
 

\subsection{4U\ 1820-30}
\begin{figure}
	\includegraphics[width=\columnwidth]{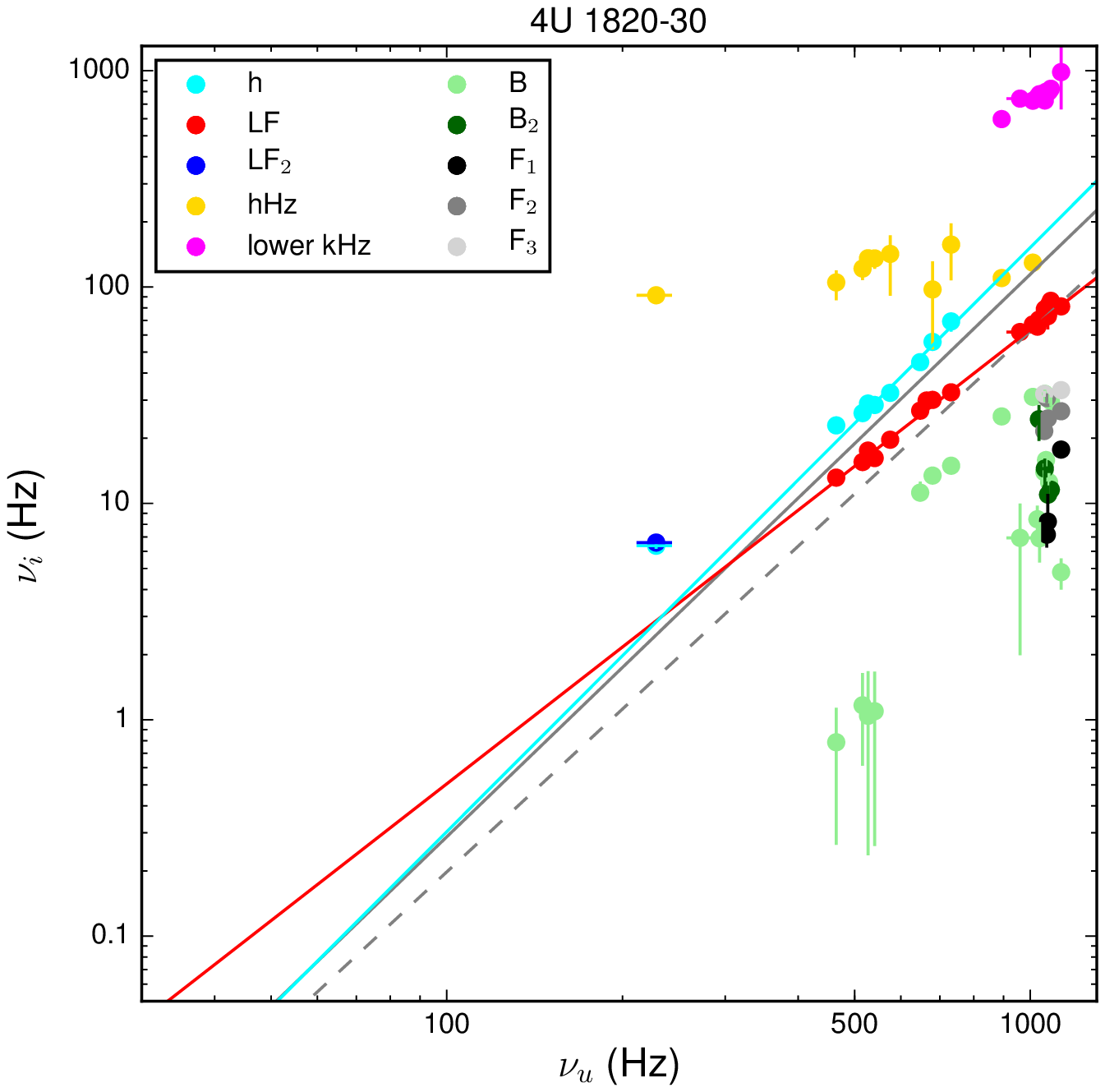}
    \caption{As in Figure \ref{fig:freqSAX}, but for 4U\ 1820--30.}
    \label{fig:freq1820}
\end{figure}

\begin{figure}
	\includegraphics[width=\columnwidth]{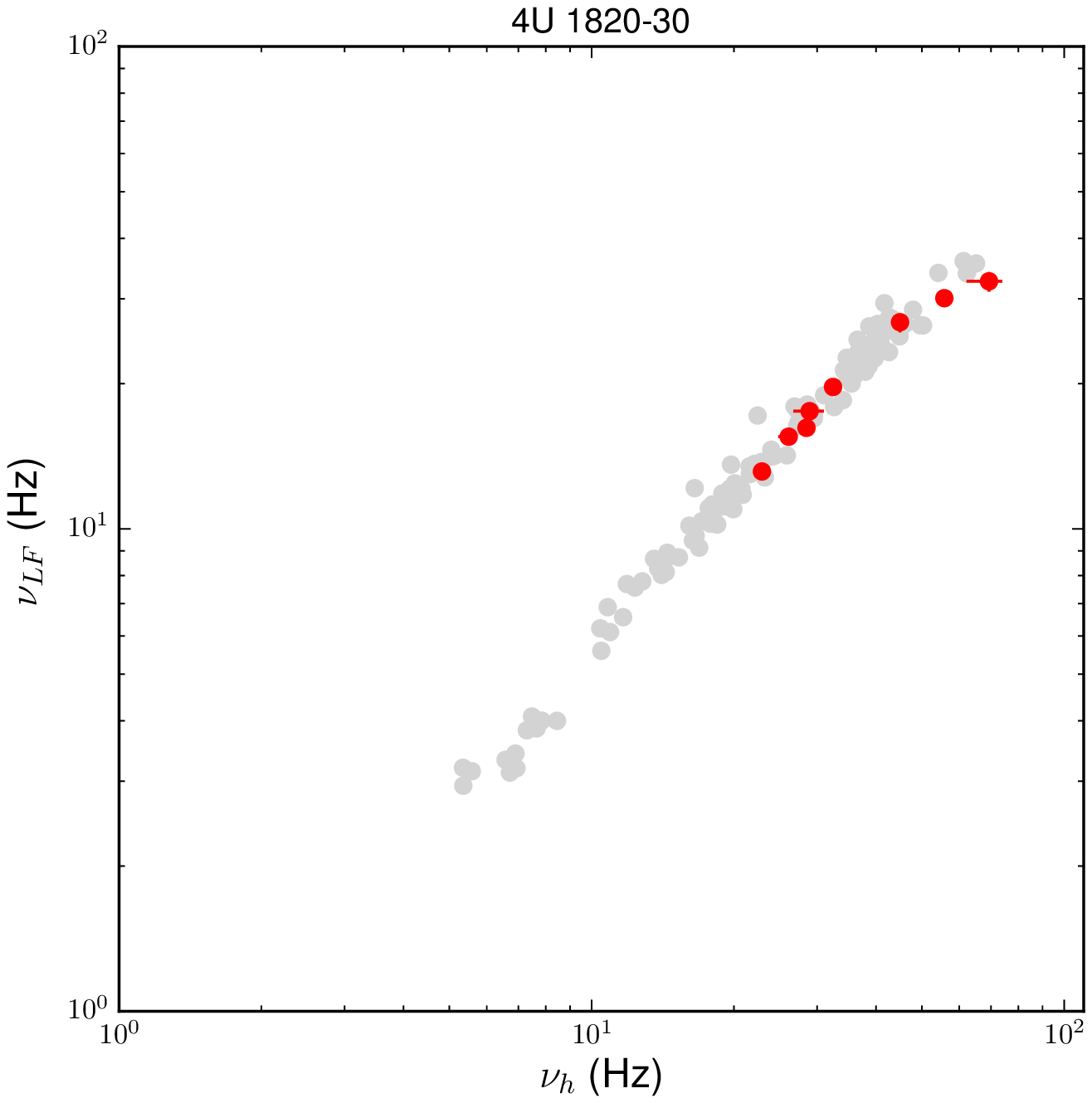}
    \caption{As in Figure \ref{fig:LFHSAX}, but for 4U\ 1820--30. }
    \label{fig:LFH1820}
\end{figure}

\begin{figure}
	\includegraphics[width=\columnwidth]{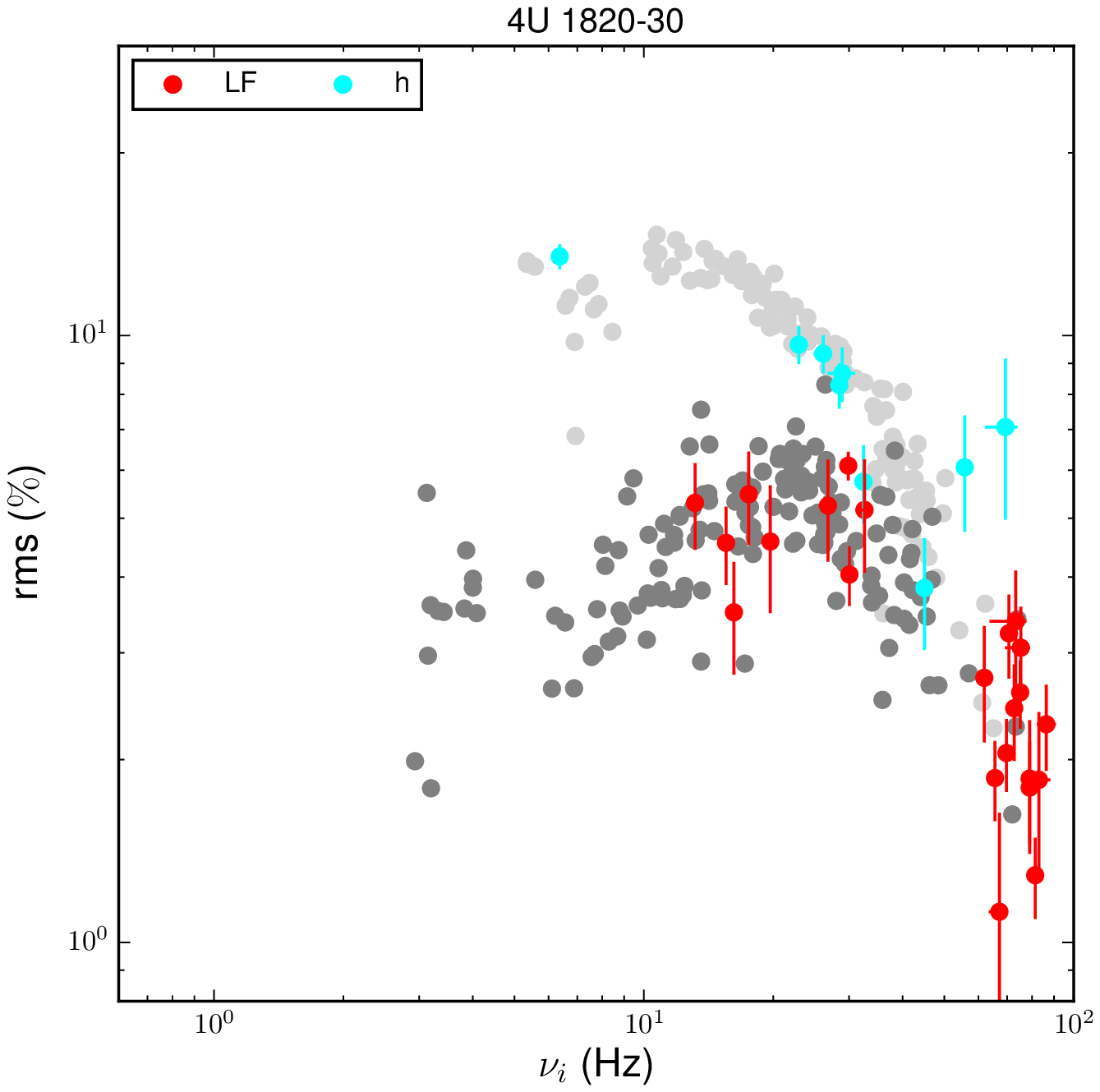}
    \caption{As in Figure \ref{fig:IPSAX}, but for 4U\ 1820--30. }
    \label{fig:IP1820}
\end{figure}

As noted earlier, in \cite{Altamirano:2005} shifted frequency correlations with respect to other atoll sources were reported for 4U\ 1820--30, similar to those seen for SAX J1808 and other pulsating sources \citep{vanStraaten:2005}. In that work, only observations taken when the source was in the island state (high hard color, $\nu_u$<700 Hz) were used. Here, we include all observations in the RXTE-archive, including those taken when the source was in the banana state (low hard color, $\nu_u$>700 Hz). As in SAX J1808, we fit asymmetric flaring features $<$10 Hz with multiple Lorentzians named L$_{F_n}$. 
We apply a 2-20 keV energy selection, after verifying that this selection optimizes the detection significance of QPOs at low and high frequency. 
We choose to fit the power spectra in the 0.9-3000 Hz domain in order to avoid fitting the noise present below 0.9 Hz. This does not affect the frequencies fitted $>$0.9 Hz. We find that the LF,h QPO show similar correlations with $\nu_u$ and each other as presented in DK17 for the non-pulsating sources, see Figures \ref{fig:freq1820}, \ref{fig:LFH1820} and \ref{fig:IP1820}.

When $\nu_u\sim$1000 Hz, only one component is fitted $<$100 Hz, which we identify as L$_{LF}$. We offer fits to two groups of data, Group 1 contains all frequencies and Group 2 (only relevant for L$_{LF}$) contains data where $\nu_u<$800 Hz and L$_{LF}$ and L$_h$ are usually simultaneously detected.

\subsection{SAX J1810.8--2609}

\begin{figure}
	\includegraphics[width=\columnwidth]{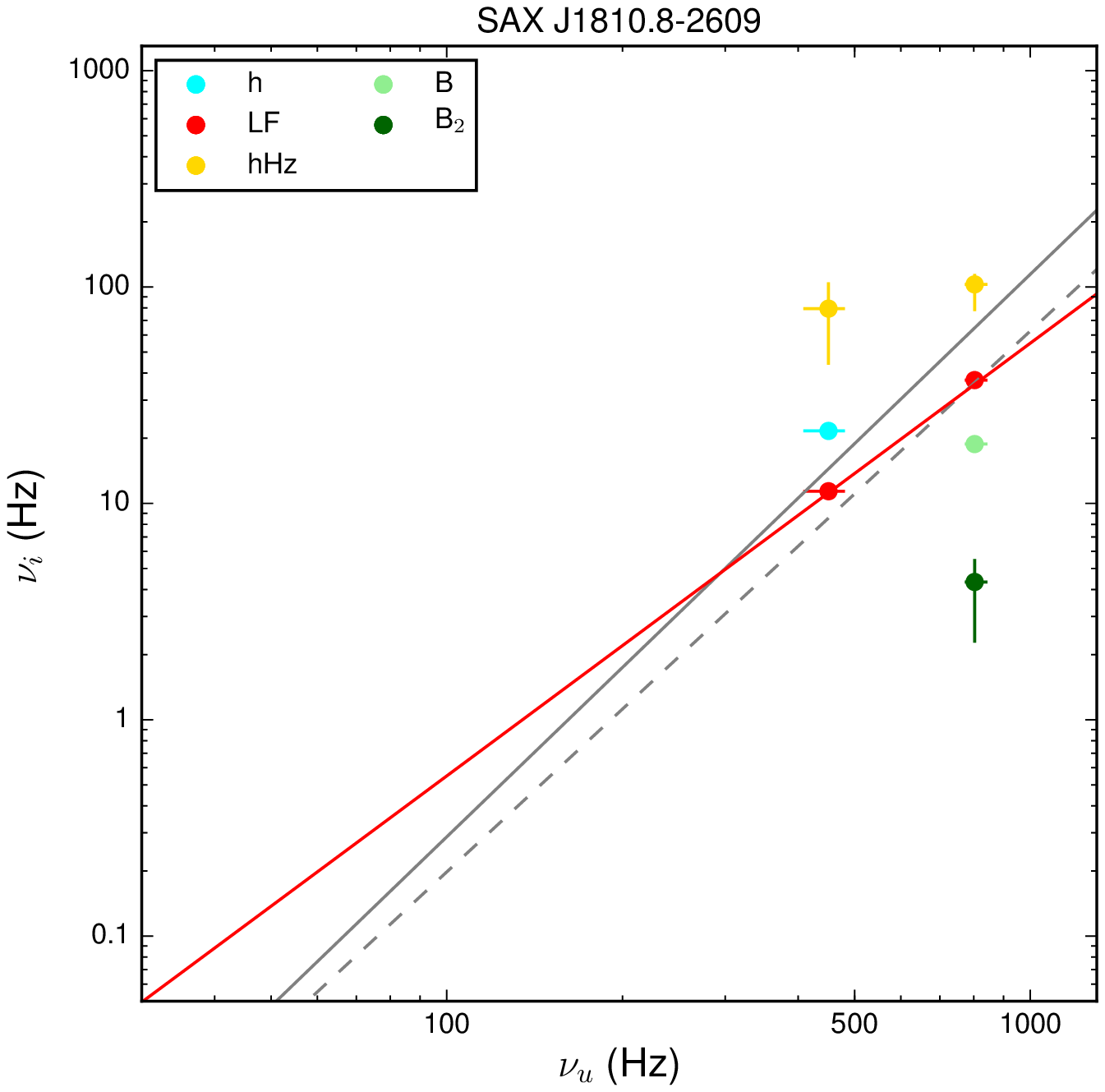}
    \caption{As in Figure \ref{fig:freqSAX}, but for SAX J1810. }
    \label{fig:freq1810}
\end{figure}

\begin{figure}
	\includegraphics[width=\columnwidth]{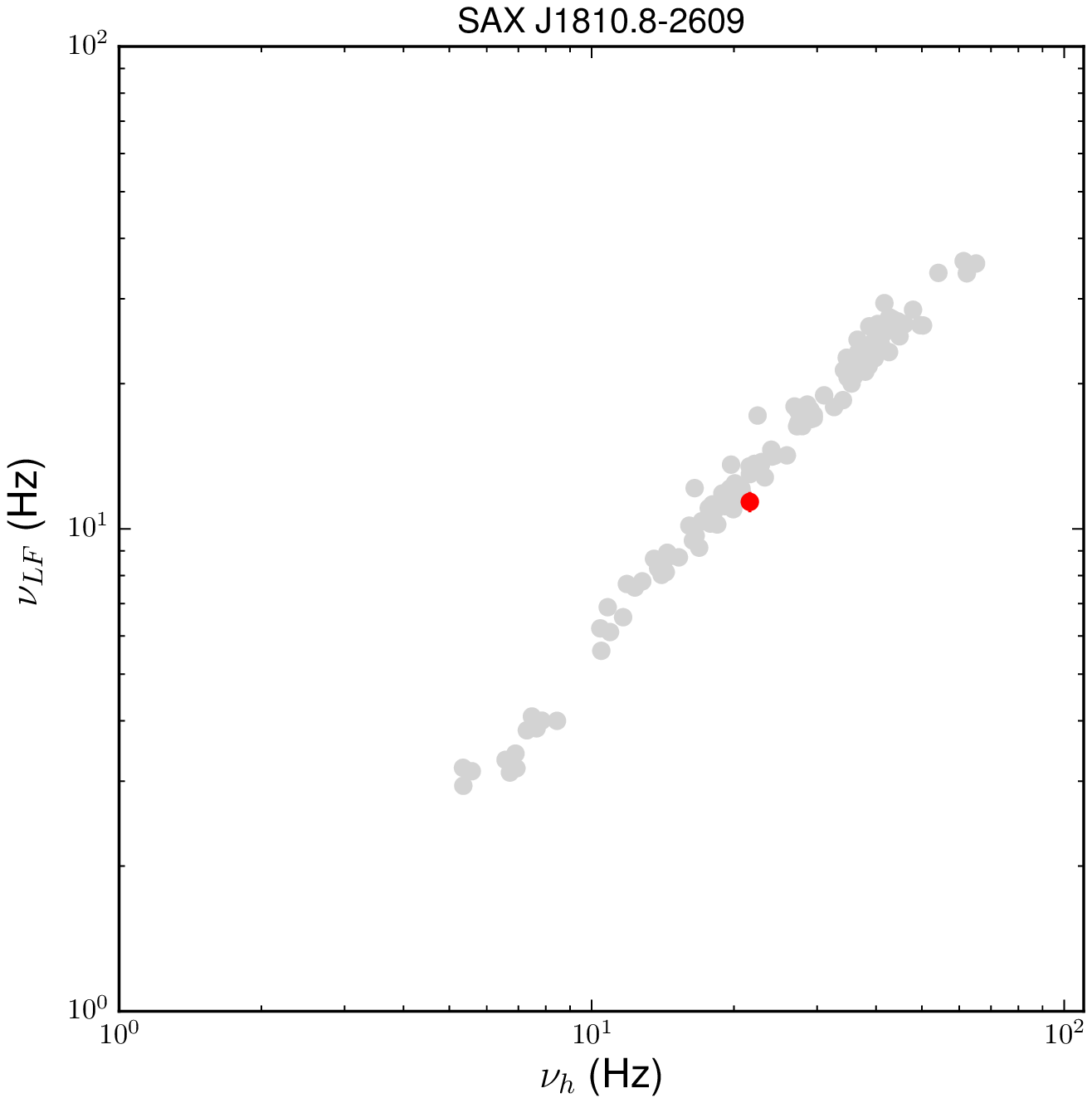}
    \caption{As in Figure \ref{fig:LFHSAX}, but for SAX J1810. }
    \label{fig:LFH1810}
\end{figure}

\begin{figure}
	\includegraphics[width=\columnwidth]{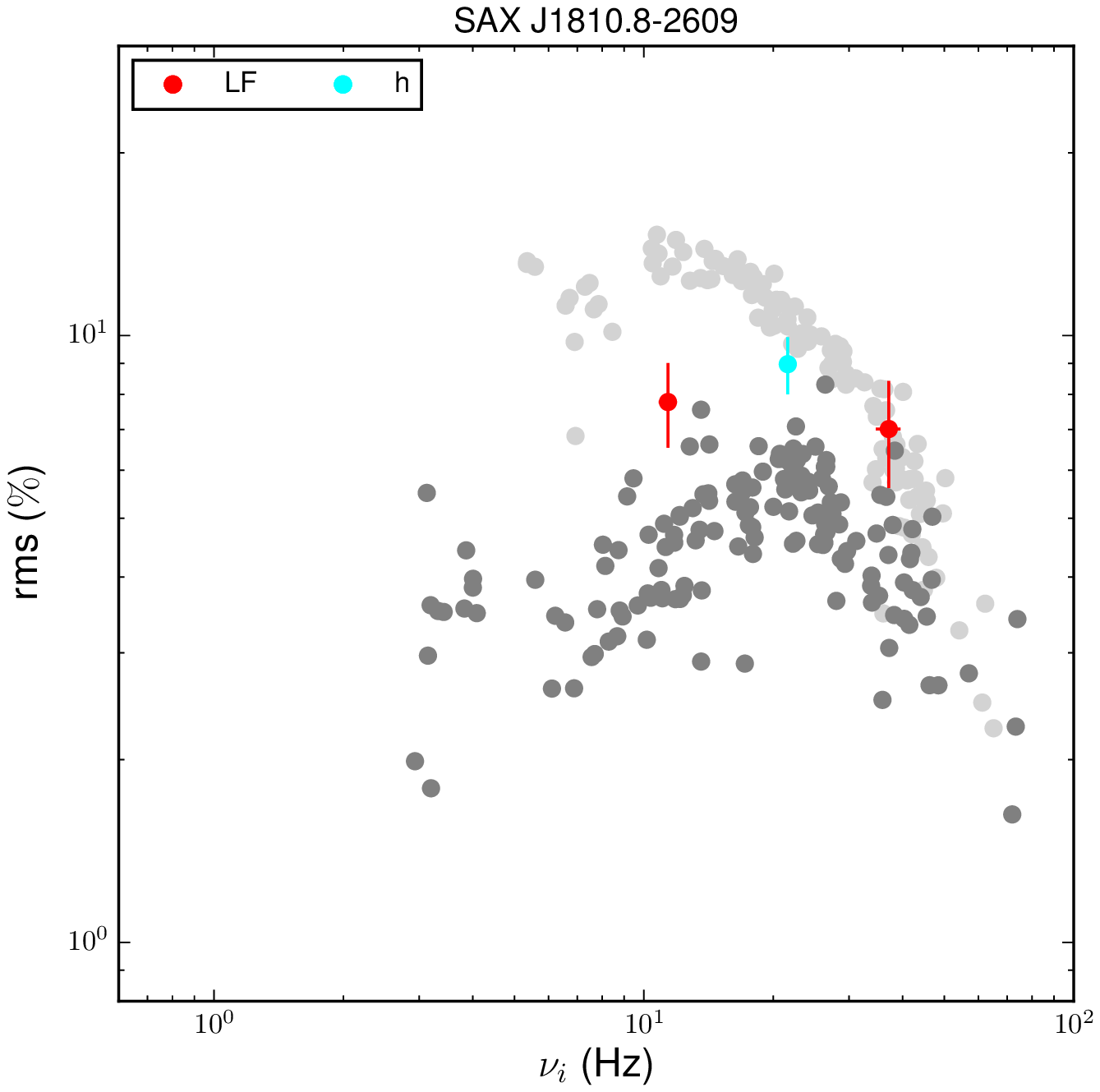}
    \caption{As in Figure \ref{fig:IPSAX}, but for SAX J1810. }
    \label{fig:IP1810}
\end{figure}

Recently, \cite{Bilous:2018} reported burst oscillations at 513.4-531.9 Hz, adding SAX J1810.8--2609 to the 'known-spin' source category. We analyze all data available in the RXTE-archive; 88 observations pass our filtering criteria. We use all energy channels to create the PDS. Due to the low signal-to-noise of these data, an energy selection worsens QPO detection significance.  We find that the source shows the canonical atoll-like aperiodic timing features with a newly discovered kHz QPO ($\nu_u\sim$400-800 Hz) and L$_{LF}$ (3.8$\sigma$) and L$_{h}$ (4.8$\sigma$) that are detected simultaneously. Unfortunately, the data are sparse. We find that $\nu_{LF,h}$ vs. $\nu_u$ roughly match the non-pulsating frequency correlations, see Figures \ref{fig:freq1810}, \ref{fig:LFH1810}, \ref{fig:IP1810}.

\subsection{Remaining AMXPs}
Pulsars XTE\ J1814-338, XTE\ J1751-305 and XTE\ J0929-314, included in the ($\nu_{max}$) timing study by \cite{vanStraaten:2005}, are observed when the source was (mostly) in the island state. Here the kHz QPOs are broad, and often zero-centered. These data are not suited for our analysis. \\
For SAX J1748.9--2021 and IGR J17511--3057 we do not fit any significant simultaneously present LF and kHz QPOs.

\subsection{Fit parameters}
 \begin{table*}
\centering
  \tabcolsep=0.1cm
 {
\scalebox{0.8}{
\small
\rotatebox{90}{
\begin{tabular}{c c c c c c c c c c c c c c c c c}

\hline	
Source & ObsID & No. of  &   & L$_{b}$ &  &  &  L$_{LF}$ &  &  & L$_h$ & &  & L$_u$ & &HC & $\chi^2/dof$ \\ 
      &       &  power spectra   & $\nu$ (Hz) & FWHM (Hz)  & IP $\times$10$^{-3}$ & $\nu$ (Hz) & FWHM (Hz) & IP $\times$10$^{-3}$ &$\nu$ (Hz)& FWHM (Hz) & IP $\times$10$^{-3}$  &$\nu$ (Hz)& FWHM (Hz) & IP $\times$10$^{-3}$  & & \\
       \hline
SAX\ J1808 &30411-01-03-00&826&0.20$^{+0.01}_{-0.02}$&0.70$^{+0.04}_{-0.04}$&33.0$^{+1.6}_{-1.7}$&-&-&-&1.38$^{+0.22}_{-0.27}$&4.94$^{+0.32}_{-0.30}$&31.1$^{+3.2}_{-3.0}$&91.6$^{+11.5}_{-15.0}$&243$^{+18}_{-16}$&40.3$^{+4.1}_{-3.7}$&1.087$\pm{0.002}$ &403.94/329\\
&30411-01-04-00&360&0.11$^{+0.03}_{-0.04}$&0.72$^{+0.06}_{-0.05}$&40.5$^{+2.5}_{-2.1}$&-&-&-&1.77$^{+0.24}_{-0.22}$&4.37$^{+0.30}_{-0.37}$&24.4$^{+2.6}_{-3.0}$&86.8$^{+20.1}_{-26.2}$&284$^{+28}_{-27}$&39.4$^{+4.7}_{-4.0}$&1.090$\pm{0.002}$ &375.02/330\\
&30411-01-05-00&890&0.12$^{+0.02}_{-0.03}$&0.71$^{+0.04}_{-0.04}$&42.4$\pm{1.8}$&-&-&-&1.69$^{+0.16}_{-0.18}$&4.02$^{+0.22}_{-0.23}$&23.5$\pm{2.2}$&86.5$^{+10.1}_{-12.5}$&242$^{+17}_{-15}$&39.8$^{+2.8}_{-2.5}$&1.089$\pm{0.002}$ &346.99/330\\
&30411-01-06-000&808&0.14$^{+0.02}_{-0.03}$&0.65$^{+0.05}_{-0.05}$&38.6$^{+2.6}_{-2.4}$&-&-&-&1.01$^{+0.29}_{-0.32}$&4.76$^{+0.22}_{-0.21}$&31.0$^{+3.2}_{-3.1}$&108$^{+18}_{-19}$&272$\pm{25}$&32.2$^{+3.3}_{-3.4}$&1.086$\pm{0.002}$ &351.57/330\\
&30411-01-06-00&645&0.12$^{+0.03}_{-0.03}$&0.72$\pm{0.05}$&44.5$^{+2.3}_{-2.5}$&-&-&-&1.68$^{+0.23}_{-0.29}$&4.18$^{+0.36}_{-0.36}$&22.3$^{+3.4}_{-3.0}$&69.6$^{+20.0}_{-28.0}$&282$^{+27}_{-24}$&42.3$^{+4.5}_{-3.8}$&0.001$\pm{0.001}$ &417.65/330\\
&30411-01-07-00&327&0.13$^{+0.04}_{-0.05}$&0.73$\pm{0.08}$&41.9$^{+3.3}_{-3.9}$&-&-&-&1.74$^{+0.41}_{-0.59}$&4.75$\pm{0.56}$&23.7$^{+5.5}_{-4.4}$&79.6$^{+31.1}_{-44.6}$&275$^{+43}_{-37}$&42.4$^{+8.0}_{-6.8}$&1.092$\pm{0.003}$ &358.05/330\\
&30411-01-08-00&1047&0.23$^{+0.03}_{-0.04}$&1.67$^{+0.09}_{-0.08}$&40.5$\pm{1.5}$&-&-&-&3.96$^{+0.49}_{-0.55}$&7.87$^{+0.86}_{-0.90}$&14.6$\pm{2.2}$&188$^{+19}_{-22}$&226$^{+50}_{-45}$&18.9$^{+4.5}_{-4.0}$&1.076$\pm{0.002}$ &348.42/330\\
&30411-01-09-00&1010&0.32$^{+0.03}_{-0.04}$&1.79$^{+0.11}_{-0.10}$&43.2$^{+2.6}_{-3.3}$&-&-&-&2.21$^{+2.65}_{-8.69}$&16.9$^{+4.4}_{-4.8}$&24.6$^{+10.9}_{-7.1}$&fixed&230$^{+73}_{-45}$&57.4$^{+3.9}_{-5.8}$&1.070$\pm{0.003}$ &356.79/333\\
&70080-01-01-000&349&5.53$^{+0.17}_{-0.19}$&16.9$^{+0.7}_{-0.6}$&31.0$^{+0.7}_{-0.6}$&36.7$^{+0.7}_{-0.8}$&13.7$^{+3.2}_{-3.8}$&4.61$^{+1.20}_{-1.59}$&55.0$^{+2.3}_{-3.5}$&32.1$^{+8.8}_{-6.1}$&7.43$^{+2.74}_{-1.65}$&596$\pm{4}$&70.7$^{+16.2}_{-12.5}$&7.30$^{+1.12}_{-1.01}$&0.978$\pm{0.002}$ &338.13/323\\
&70080-01-01-01&1009&15.1$\pm{0.4}$&10.5$^{+1.4}_{-1.3}$&6.93$^{+1.97}_{-1.53}$&46.0$\pm{0.6}$&18.6$^{+2.3}_{-2.0}$&4.92$^{+0.60}_{-0.54}$&74.2$^{+1.1}_{-1.2}$&27.9$^{+4.5}_{-3.9}$&4.71$^{+0.72}_{-0.65}$&685$\pm{4}$&93.1$^{+16.6}_{-13.9}$&5.80$^{+0.86}_{-0.76}$&0.974$\pm{0.001}$ &268.09/323\\
&70080-01-01-020&1165&3.65$^{+0.08}_{-0.09}$&13.1$^{+0.3}_{-0.3}$&29.2$\pm{0.4}$&27.8$\pm{0.5}$&10.62$^{+1.47}_{-1.56}$&4.55$^{+1.11}_{-1.14}$&38.0$^{+1.0}_{-1.1}$&22.1$^{+2.4}_{-2.3}$&11.0$^{+2.0}_{-1.8}$&491$\pm{7}$&172$^{+21}_{-18}$&10.7$^{+1.0}_{-0.9}$&0.992$\pm{0.001}$ &407.85/326\\
&70080-01-01-03&323&14.5$^{+0.5}_{-0.7}$&10.7$\pm{2.0}$&8.25$^{+2.92}_{-2.57}$&46.2$^{+1.3}_{-1.2}$&18.7$^{+4.9}_{-3.9}$&4.56$^{+1.06}_{-0.94}$&76.4$^{+1.8}_{-2.1}$&27.1$^{+8.1}_{-6.2}$&5.26$^{+1.45}_{-1.20}$&702$\pm{7}$&85.9$^{+21.9}_{-17.8}$&6.56$^{+1.16}_{-1.09}$&0.975$\pm{0.003}$ &295.84/323\\
&70080-01-02-00&1446&1.23$^{+0.05}_{-0.05}$&6.04$^{+0.14}_{-0.13}$&24.5$^{+0.4}_{-0.3}$&10.6$\pm{0.1}$&1.97$^{+0.80}_{-0.63}$&0.79$^{+0.34}_{-0.24}$&16.5$^{+0.3}_{-0.3}$&15.7$^{+0.8}_{-0.9}$&19.1$^{+1.2}_{-1.3}$&336$\pm{6}$&164$^{+21}_{-18}$&10.8$^{+1.4}_{-1.2}$&0.001$\pm{0.002}$ &398.87/326\\
&70080-01-02-02&1242&1.39$^{+0.05}_{-0.05}$&5.67$^{+0.15}_{-0.14}$&23.2$^{+0.4}_{-0.4}$&10.3$\pm{0.3}$&3.31$^{+1.05}_{-0.85}$&1.15$^{+0.46}_{-0.35}$&17.1$\pm{0.3}$&15.6$\pm{1.1}$&18.2$^{+1.5
}_{-1.6}$&330$\pm{5}$&141$^{+25}_{-18}$&10.5$^{+1.7}_{-1.3}$&1.022$\pm{0.005}$ &395.46/326\\
&70080-01-02-05&1267&1.42$^{+0.05}_{-0.05}$&5.42$^{+0.13}_{-0.14}$&23.8$^{+0.4}_{-0.4}$&10.4$^{+0.4}_{-0.5}$&3.22$^{+1.19}_{-1.13}$&0.80$^{+0.40}_{-0.35}$&16.7$\pm{0.3}$&16.0$^{+0.9}_{-0.7}$&20.3$^{+1.2}_{-1.0}$&326$^{+6}_{-5}$&120$^{+23}_{-22}$&8.69$^{+1.42}_{-1.49}$&1.023$\pm{0.003}$ &367.66/327\\
&70080-01-02-08&1147&0.85$^{+0.05}_{-0.05}$&4.25$^{+0.13}_{-0.15}$&26.3$^{+0.6}_{-0.7}$&-&-&-&11.0$^{+0.3}_{-0.4}$&16.6$^{+1.2}_{-1.0}$&19.6$^{+1.6}_{-1.3}$&285$^{+7}_{-8}$&135$^{+38}_{-27}$&10.1$^{+2.9}_{-2.0}$&1.040$\pm{0.001}$ &333.35/329\\
&70080-01-02-09&806&2.30$^{+0.09}_{-0.11}$&8.99$^{+0.35}_{-0.29}$&24.9$\pm{0.6}$&-&-&-&23.4$\pm{0.3}$&20.0$^{+1.4}_{-1.5}$&20.2$^{+1.5}_{-1.9}$&395$^{+5}_{-4}$&85.0$^{+10.4}_{-10.0}$&10.0$\pm{1.0}$&1.021$\pm{0.002}$ &432.29/329\\
&70080-01-02-11&942&1.79$^{+0.10}_{-0.10}$&7.83$^{+0.32}_{-0.32}$&24.2$\pm{0.6}$&13.7$^{+0.7}_{-0.8}$&5.43$^{+2.81}_{-2.67}$&1.46$^{+1.04}_{-0.81}$&21.7$\pm{0.5}$&16.7$^{+1.6}_{-1.4}$&18.4$^{+2.0}_{-1.8}$&364$\pm{9}$&152$^{+26}_{-22}$&12.2$^{+1.8}_{-1.7}$&1.025$\pm{0.002}$ &364.00/327\\
&70080-01-02-12&415&1.38$^{+0.15}_{-0.19}$&6.73$^{+0.55}_{-0.47}$&23.0$^{+1.1}_{-1.0}$&-&-&-&17.7$\pm{0.5}$&15.2$^{+2.3}_{-2.6}$&18.4$^{+2.8}_{-3.6}$&333$^{+14}_{-17}$&118$^{+46}_{-33}$&10.1$^{+3.1}_{-2.5}$&1.030$\pm{0.005}$ &359.38/330\\
&70080-01-02-13&1249&1.54$^{+0.08}_{-0.09}$&6.90$^{+0.29}_{-0.24}$&24.7$\pm{0.6}$&-&-&-&18.2$\pm{+0.3}$&17.4$^{+1.2}_{-1.3}$&22.3$^{+1.5}_{-1.9}$&348$^{+13}_{-10}$&154$^{+31}_{-28}$&11.2$^{+1.8}_{-2.0}$&1.041$\pm{0.005}$ &380.46/329\\
&70080-01-02-17&467&1.41$^{+0.16}_{-0.17}$&6.65$^{+0.53}_{-0.51}$&23.2$^{+1.2}_{-1.3}$&-&-&-&19.2$^{+0.7}_{-0.8}$&24.8$^{+2.8}_{-2.5}$&28.5$\pm{2.6}$&326$^{+19}_{-20}$&223$^{+66}_{-51}$&18.7$^{+3.5}_{-3.2}$&1.031$\pm{0.005}$ &326.55/329\\
&70080-01-02-19&320&2.21$^{+0.16}_{-0.20}$&9.13$^{+0.63}_{-0.49}$&24.7$^{+1.0}_{-0.8}$&-&-&-&23.5$^{+0.6}_{-0.5}$&19.3$^{+2.2}_{-2.3}$&18.2$^{+2.0}_{-2.4}$&391$^{+12}_{-13}$&114$^{+43}_{-30}$&9.16$^{+2.65}_{-2.25}$&1.019$\pm{0.003}$ &332.45/330\\
&70080-01-02-20&138&2.26$^{+0.21}_{-0.24}$&8.29$^{+0.64}_{-0.58}$&24.3$\pm{1.1}$&-&-&-&22.4$\pm{0.6}$&19.4$^{+3.3}_{-3.2}$&19.4$^{+3.3}_{-3.6}$&396$^{+11}_{-14}$&101$^{+64}_{-46}$&10.3$^{+4.1}_{-3.9}$&1.022$\pm{0.004}$ &274.34/330\\
&70080-01-03-000&1030&1.42$^{+0.09}_{-0.10}$&6.38$^{+0.27}_{-0.27}$&24.7$^{+0.6}_{-0.7}$&-&-&-&18.5$\pm{0.4}$&19.9$^{+1.6}_{-1.7}$&23.5$^{+1.9}_{-2.4}$&324$^{+12}_{-11}$&123$^{+65}_{-59}$&8.42$^{+4.32}_{-3.76}$&1.037$\pm{0.002}$ &308.96/329\\
&70080-01-03-01&1366&1.27$^{+0.07}_{-0.07}$&5.86$^{+0.19}_{-0.20}$&26.4$^{+0.5}_{-0.6}$&-&-&-&17.0$\pm{0.3}$&17.1$^{+1.1}_{-1.0}$&21.2$^{+1.3}_{-1.2}$&315$^{+12}_{-15}$&166$^{+55}_{-52}$&11.4$^{+3.7}_{-3.5}$&0.001$\pm{0.003}$ &363.45/330\\
&70080-02-01-000&1326&1.99$^{+0.30}_{-0.38}$&8.64$^{+1.00}_{-0.86}$&23.4$^{+1.5}_{-1.4}$&-&-&-&26.8$\pm{1.05}$&19.7$^{+3.9}_{-3.3}$&18.9$^{+3.1}_{-3.0}$&400$^{+12}_{-11}$&55$^{+34}_{-21}$&9.03$^{+4.34}_{-3.51}$&1.049$\pm{0.004}$ &321.10/330\\
&70080-03-04-00&116&9.06$^{+0.63}_{-0.70}$&10.1$^{+2.2}_{-2.6}$&13.0$^{+5.2}_{-5.0}$&-&-&-&33.4$^{+3.6}_{-4.0}$&59.1$^{+8.1}_{-7.3}$&26.1$^{+3.0}_{-4.5}$&564$^{+14}_{-11}$&71.3$^{+21.8}_{-16.7}$&6.38$^{+1.47}_{-1.37}$&0.985$\pm{0.003}$ &341.90/330\\
&91056-01-01-02&1336&1.61$^{+0.05}_{-0.06}$&5.50$^{+0.18}_{-0.17}$&23.8$^{+0.4}_{-0.5}$&-&-&-&17.3$\pm{0.2}$&15.4$^{+1.0}_{-0.9}$&20.4$\pm{1.1}$&309$\pm{10}$&108$^{+45}_{-34}$&6.60$^{+2.39}_{-1.98}$&1.039$\pm{0.006}$ &345.89/330\\
&91056-01-02-01&1013&2.90$^{+0.11}_{-0.11}$&9.26$^{+0.32}_{-0.31}$&21.1$\pm{0.5}$&-&-&-&26.7$\pm{0.3}$&18.4$^{+1.4}_{-1.3}$&14.3$^{+1.2}_{-1.1}$&402$^{+4.5}_{-4.7}$&67.8$^{+14.5}_{-12.4}$&6.76$^{+0.98}_{-0.94}$&1.022$\pm{0.002}$ &384.21/330\\
&91056-01-02-02&805&3.81$^{+0.12}_{-0.13}$&13.1$\pm{0.4}$&27.7$\pm{0.5}$&-&-&-&36.3$^{+0.5}_{-0.6}$&28.0$^{+2.2}_{-1.9}$&14.8$^{+1.2}_{-1.1}$&519$\pm{6}$&91.2$^{+22.1}_{-18.3}$&6.77$^{+1.24}_{-1.21}$&0.994$\pm{0.002}$ &347.33/330\\
&91056-01-02-030&1004&3.87$^{+0.11}_{-0.11}$&13.5$^{+0.4}_{-0.3}$&29.3$^{+0.5}_{-0.6}$&-&-&-&35.8$^{+0.5}_{-0.4}$&26.4$^{+1.9}_{-2.0}$&14.5$^{+1.2}_{-1.4}$&531$\pm{5}$&110$^{+19}_{-18}$&9.07$^{+1.08}_{-1.15}$&0.990$\pm{0.002}$ &383.10/329\\
&91056-01-02-03&170&3.92$^{+0.18}_{-0.19}$&11.2$\pm{0.5}$&28.8$\pm{0.8}$&-&-&-&33.2$^{+0.7}_{-0.6}$&23.1$^{+2.9}_{-2.7}$&17.0$^{+1.9}_{-2.2}$&509$^{+53}_{-44}$&271$^{+316}_{-283}$&9.52$^{+5.5}_{-5.2}$&0.001$\pm{0.002}$ &358.73/329\\
&91056-01-02-04&2404&14.4$^{+0.5}_{-0.6}$&11.6$\pm{1.6}$&6.61$^{+2.1}_{-1.7}$&47.4$\pm{0.5}$&13.8$^{+3.0}_{-2.6}$&3.23$^{+1.06}_{-0.78}$&70.6$^{+5.3}_{-5.4}$&70.4$^{+18.0}_{-29.0}$&9.84$^{+2.95}_{-4.73}$&668$^{+16}_{-23}$&156$^{+89}_{-45}$&4.57$^{+2.78}_{-1.45}$&0.002$\pm{0.059}$ &0.94312230\\
&91056-01-02-08&786&3.42$^{+0.14}_{-0.14}$&12.8$\pm{0.4}$&29.4$\pm{0.6}$&-&-&-&35.0$\pm{0.4}$&21.5$^{+1.4}_{-1.3}$&16.3$\pm{1.1}$&503$^{+6}_{-7}$&98.3$^{+30.4}_{-21.6}$&8.11$^{+1.79}_{-1.51}$&0.996$\pm{0.002}$ &350.56/329\\
&91056-01-03-00&1585&7.40$^{+0.55}_{-0.68}$&7.98$^{+0.97}_{-1.05}$&10.3$^{+4.5}_{-3.2}$&-&-&-&34.2$\pm{0.3}$&26.2$\pm{1.4}$&19.7$^{+1.3}_{-1.3}$&489$^{+8}_{-10}$&170$^{+40}_{-31}$&11.0$\pm{0.1}$&1.003$\pm{0.002}$ &344.34/326\\
&91056-01-03-04&1420&2.70$\pm{0.17}$&12.7$\pm{0.4}$&25.9$\pm{+0.6}$&-&-&-&33.3$\pm{0.3}$&19.8$^{+1.5}_{-1.3}$&16.1$\pm{1.1}$&485$^{+18}_{-16}$&184$\pm{36}$&8.43$^{+4.02}_{-3.34}$&1.017$\pm{0.003}$ &407.71/330\\
&91056-01-03-06&2510&1.53$^{+0.15}_{-0.16}$&9.92$^{+0.45}_{-0.44}$&25.8$\pm{0.8}$&-&-&-&25.3$^{+0.5}_{-0.6}$&24.2$\pm{2.9}$&18.1$^{+2.4}_{-2.7}$&381$^{+21}_{-20}$&165$^{+88}_{-66}$&7.13$^{+3.17}_{-2.77}$&1.024$\pm{0.002}$ &360.40/330\\
&93027-01-01-03&1327&1.35$^{+0.11}_{-0.13}$&6.66$^{+0.35}_{-0.32}$&24.5$^{+0.8}_{-0.9}$&-&-&-&18.2$\pm{0.3}$&15.5$^{+1.6}_{-1.8}$&20.1$^{+2.3}_{-3.0}$&366$^{+18}_{-16}$&163$^{+51}_{-41}$&11.2$^{+3.1}_{-2.9}$&1.030$\pm{0.006}$ &377.26/329\\
&93027-01-01-04&1407&4.24$^{+0.53}_{-0.35}$&6.70$^{+0.61}_{-0.79}$&17.0$^{+2.3}_{-4.8}$&-&-&-&23.0$^{+0.5}_{-0.6}$&25.1$^{+2.1}_{-1.9}$&26.1$\pm{1.9}$&414$\pm{3}$&33.3$^{+14.7}_{-12.0}$&5.64$^{+1.84}_{-1.56}$&1.009$\pm{0.008}$ &330.48/327\\
&93027-01-02-02&1171&-&-&-&-&-&-&34.8$^{+1.3}_{-1.5}$&45.9$^{+3.9}_{-3.6}$&41.8$^{+2.5}_{-2.4}$&578$^{+34}_{-38}$&279$^{+85}_{-66}$&18.8$^{+3.7}_{-3.5}$&0.999$\pm{0.006}$ &361.96/329\\
&93027-01-02-03&1112&4.61$^{+0.26}_{-0.28}$&16.1$\pm{1.2}$&24.2$^{+1.3}_{-1.4}$&-&-&-&40.2$^{+2.2}_{-2.7}$&49.4$^{+8.1}_{-6.6}$&18.0$^{+2.5}_{-2.2}$&436$^{+57}_{-105}$&727$^{+357}_{-211}$&26.9$^{+5.5}_{-4.5}$&0.980$\pm{0.004}$ &372.35/332\\
&96027-01-01-00&970&-&-&-&47.1$\pm{0.5}$&32.6$\pm{2.6}$&24.5$^{+1.8}_{-2.1}$&-&-&-&657$\pm{4}$&92.9$^{+11.0}_{-9.9}$&14.5$^{+1.4}_{-1.3}$&0.987$\pm{0.002}$ &567.74/326\\
&96027-01-01-03&756&-&-&-&51.0$^{+0.6}_{-0.7}$&10.3$^{+2.9}_{-2.3}$&4.97$^{+1.74}_{-1.35}$&51.1$^{+3.0}_{-2.9}$&68.1$^{+10.6}_{-8.6}$&25.0$^{+2.2}_{-2.7}$&716$\pm{4}$&104$^{+13}_{-12}$&19.6$\pm{1.6}$&0.976$\pm{0.003}$ &498.05/323\\
&96027-01-01-05&605&-&-&-&52.07$^{+0.70}_{-0.70}$&22.8$^{+3.3}_{-3.0}$&15.0$^{+1.8}_{-1.9}$&-&-&-&711$\pm{4}$&82.8$^{+14.0}_{-12.4}$&17.2$\pm{2.1}$&0.965$\pm{0.005}$ &459.47/327\\
&96027-01-01-07&1162&-&-&-&50.2$^{+1.9}_{-2.5}$&20.8$^{+7.9}_{-11.0}$&4.89$^{+1.67}_{-3.14}$&78.6$^{+3.5}_{-14.2}$&27.2$^{+31.4}_{-13.5}$&3.26$^{+5.04}_{-1.58}$&759$^{+16}_{-15}$&147$^{+49}_{-37}$&7.80$^{+2.12}_{-1.83}$&0.001$\pm{0.002}$ &345.78/248\\
&96027-01-01-08&880&2.56$^{+0.38}_{-0.45}$&13.1$^{+1.2}_{-1.1}$&30.4$^{+1.7}_{-1.6}$&-&-&-&29.2$\pm{0.9}$&17.6$^{+3.8}_{-3.1}$&14.5$^{+2.8}_{-2.6}$&476$^{+27}_{-25}$&173$^{+76}_{-62}$&16.0$^{+5.7}_{-6.0}$&1.002$\pm{0.005}$ &340.00/330\\
 	 \hline		
SAX J1810 &93044-02-07-00&187&18.8$^{+0.5}_{-0.6}$&5.92$^{+2.15}_{-1.54}$&5.20$^{+1.88}_{-1.59}$&37.1$^{+2.4}_{-2.5}$&16.6$^{+8.5}_{-5.7}$&4.92$^{+2.09}_{-1.85}$&-&-&-&803$^{+41}_{-31}$&221$^{+127}_{-98}$&14.8$^{+4.8}_{-4.6}$&0.764$\pm{0.003}$ &297.53/326\\
&93093-01-01-00&1442&0.25$^{+0.24}_{-0.34}$&7.00$^{+0.61}_{-0.53}$&23.0$^{+1.4}_{-1.3}$&11.4$^{+0.5}_{-0.6}$&8.40$^{+2.55}_{-1.95}$&6.04$^{+2.16}_{-1.69}$&21.6$^{+0.8}_{-0.9}$&13.0$^{+2.7}_{-2.3}$&8.05$^{+1.79}_{-1.68}$&451$^{+30}_{-42}$&392$^{+174}_{-110}$&21.2$^{+5.4}_{-4.1}$&0.977$\pm{0.002}$ &304.63/327\\
\hline
\end{tabular}

}}}

\label{tab:pars}
\end{table*}
\newpage
 \begin{table}
\centering
  \tabcolsep=0.1cm
 {
\scalebox{0.8}{
\small
\rotatebox{90}{
\begin{tabular}{c c c c c c c c c c c c c c c c c}

\hline	
Source & ObsID & No. of  &   & L$_{b}$ &  &  &  L$_{LF}$ &  &  & L$_h$ & &  & L$_u$ & &HC & $\chi^2/dof$ \\ 
      &       &  power spectra   & $\nu$ (Hz) & FWHM (Hz)  & IP $\times$10$^{-3}$ & $\nu$ (Hz) & FWHM (Hz) & IP $\times$10$^{-3}$ &$\nu$ (Hz)& FWHM (Hz) & IP $\times$10$^{-3}$  &$\nu$ (Hz)& FWHM (Hz) & IP $\times$10$^{-3}$ & & \\
       \hline
XTE\ J1807&70134-09-02-00&1379&1.92$^{+0.99}_{-1.69}$&17.4$^{+3.8}_{-2.5}$&17.2$^{+3.8}_{-3.2}$&27.4$\pm{0.6}$&10.2$^{+3.9}_{-2.8}$&7.2$^{+3.0}_{-2.0}$&fixed &132$^{+49}_{-25}$&44.1$^{+4.1}_{-4.3}$&464$\pm{2}$&66.6$^{+5.7}_{-5.4}$&37.2$\pm{2.4}$&1.063$\pm{0.004}$ &372.40/327\\
&80145-01-02-00&1018&fixed &10.1$\pm{1.2}$&17.8$^{+2.0}_{-2.3}$&14.5$^{+0.6}_{-0.7}$&12.8$^{+4.0}_{-2.5}$&14.3$^{+4.1}_{-2.8}$&27.3$^{+0.6}_{-0.5}$&4.36$^{+4.44}_{-2.06}$&2.63$^{+1.97}_{-1.13}$&342$\pm{6}$&102$^{+21}_{-18}$&24.7${4.0}$&1.078$\pm{0.004}$ &332.02/327\\
&80145-01-03-00&1247&3.75$^{+0.67}_{-0.89}$&19.9$^{+2.5}_{-2.0}$&23.9$\pm{1.9}$&38.9$\pm{1.1}$&23.1$^{+4.6}_{-4.1}$&14.3$^{+2.9}_{-2.7}$&- &- &- &563$\pm{3}$&76.4$^{+10.9}_{-9.3}$&27.6$^{+3.2}_{-2.9}$&1.045$\pm{0.003}$ &284.28/326\\
&80145-01-03-03&96&3.46$^{+0.54}_{-0.76}$&8.20$^{+2.76}_{-6.66}$&14.3$^{+2.7}_{-20.2}$&17.0$\pm{0.5}$&5.98$^{+1.74}_{-2.57}$&10.7$^{+2.6}_{-5.6}$&30.6$^{+1.0}_{-0.9}$&5.51$^{+4.14}_{--1.00}$&5.81$^{+3.37}_{-3.65}$&371$^{+5}_{-7}$&35.9$^{+17.3}_{-12.3}$&14.6$^{+4.4}_{-3.7}$&1.071$\pm{0.007}$ &330.68/323\\
&80145-01-01-03&130&fixed &12.5$^{+2.5}_{-2.0}$&15.3$^{+2.4}_{-2.1}$&17.4$\pm{0.6}$&8.74$^{+2.13}_{-1.73}$&11.3$^{+2.4}_{-2.2}$&34.3$^{+2.7}_{-3.0}$&13.5$^{+11.9}_{-6.2}$&4.53$^{+3.25}_{-2.02}$&370$\pm{3}$&49.8$^{+9.0}_{-7.6}$&23.6$^{+3.3}_{-3.2}$&1.074$\pm{0.005}$ &321.98/328\\
&80145-01-01-04&129&fixed &16.8$^{+4.9}_{-3.4}$&15.4$^{+3.7}_{-2.8}$&17.1$\pm{0.6}$&7.21$^{+2.90}_{-2.40}$&7.96$^{+2.92}_{-2.58}$&31.9$^{+1.7}_{-1.8}$&14.6$^{+7.7}_{-4.9}$&7.89$^{+3.33}_{-2.35}$&376$\pm{3}$&55.4$^{+9.0}_{-7.8}$&29.9$^{+3.9}_{-3.7}$&1.077$\pm{0.005}$ &356.73/328\\
&80145-01-01-00&592&1.82$^{+0.59}_{-0.98}$&10.7$^{+2.6}_{-2.2}$&11.0$^{+2.7}_{-3.0}$&16.2$\pm{0.4}$&5.77$^{+2.61}_{-1.79}$&4.91$^{+2.66}_{-1.68}$&19.0$^{+7.0}_{-7.4}$&61.8$^{+8.4}_{-6.8}$&33.9$^{+5.7}_{-5.3}$&375$\pm{2}$&61.1$^{+5.8}_{-5.3}$&29.3$\pm{1.9}$&1.073$\pm{0.003}$ &354.30/326\\
&80145-01-01-01&141&fixed &226$^{+93}_{-68}$&40.7$^{+5.1}_{-4.4}$&21.2$^{+0.7}_{-0.6}$&9.79$^{+3.42}_{-2.57}$&9.92$^{+2.94}_{-2.49}$&fixed &16.3$^{+3.4}_{-2.5}$&13.0$^{+2.5}_{-2.2}$&407$\pm{3}$&65.7$^{+10.6}_{-9.6}$&31.9$\pm{3.7}$&1.067$\pm{0.005}$ &324.45/331\\
&80145-01-03-02&409&fixed &42.7$^{+4.7}_{-5.8}$&38.5$^{+4.3}_{-5.9}$&19.9$\pm{0.6}$&4.41$^{+1.71}_{-1.21}$&3.00$^{+1.17}_{-0.88}$&- &- &- &387$\pm{3}$&48.3$^{+7.3}_{-6.3}$&21.9$^{+2.9}_{-2.7}$&1.053$\pm{0.003}$ &327.98/330\\
&80145-01-02-06&1013&fixed &37.7$^{+4.9}_{-11.4}$&43.8$^{+4.6}_{-12.2}$&25.4$\pm{1.0}$&7.46$^{+6.12}_{-2.57}$&4.50$^{+5.65}_{-1.77}$&- &- &- &450$\pm{3}$&44.7$^{+9.4}_{-8.6}$&20.2$^{+3.3}_{-3.5}$&1.048$\pm{0.004}$ &368.70/330\\
&80145-01-03-01&706&fixed &28.7$^{+6.6}_{-4.9}$&33.3$^{+7.0}_{-5.9}$&30.8$^{+1.7}_{-2.1}$&22.5$^{+9.1}_{-7.3}$&15.4$^{+6.6}_{-5.7}$&- &- &- &486$^{+3.4}_{-3.2}$&53.7$^{+10.5}_{-9.1}$&25.7$^{+3.9}_{-3.8}$&1.034$\pm{0.004}$ &302.90/330\\
&80145-01-02-03&722&2.74$^{+0.36}_{-0.46}$&9.17$^{+1.55}_{-1.29}$&13.5$^{+1.8}_{-2.0}$&14.6$\pm{0.2}$&2.44$^{+1.14}_{-0.75}$&3.05$^{+1.30}_{-0.90}$&21.0$^{+2.4}_{-2.5}$&31.3$^{+5.8}_{-4.7}$&24.0$^{+4.5}_{-4.0}$&355$\pm{3}$&56.1$^{+10.8}_{-9.1}$&23.0$^{+3.7}_{-3.5}$&1.053$\pm{0.007}$ &331.42/326\\
 	 \hline		
HETE J1900.1 &91015-01-04-04&49&fixed&8.55$^{+0.47}_{-0.50}$&34.4$^{+1.5}_{-1.8}$&13.4$\pm{0.9}$&6.92$^{+5.02}_{-2.28}$&5.06$^{+3.92}_{-1.86}$&19.9$\pm{0.7}$&6.01$^{+2.61}_{-2.09}$&4.93$^{+2.21}_{-2.17}$&432$^{+19}_{-23}$&157$^{+61}_{-53}$&17.1$^{+5.2}_{-5.3}$&1.040$\pm{0.003}$ &427.29/418\\
&91015-01-04-06&50&0.97$^{+0.24}_{-0.30}$&7.00$^{+0.76}_{-0.67}$&28.7$\pm{1.9}$&10.5$^{+0.4}_{-0.5}$&3.57$^{+2.04}_{-1.33}$&3.41$^{+2.12}_{-1.43}$&19.5$^{+1.6}_{-2.2}$&20.1$^{+5.4}_{-4.3}$&17.3$^{+4.2}_{-3.6}$&493$^{+15}_{-18}$&157$^{+54}_{-41}$&23.4$^{+5.1}_{-4.9}$&1.032$\pm{0.004}$ &409.19/417\\
&91015-01-05-00&48&0.75$^{+0.73}_{-1.21}$&14.7$^{+2.2}_{-1.7}$&33.1$^{+3.3}_{-2.6}$&17.5$\pm{0.5}$&3.95$^{+2.28}_{-1.53}$&3.20$^{+1.67}_{-1.26}$&27.8$^{+1.5}_{-2.0}$&15.0$^{+8.6}_{-6.0}$&8.04$^{+4.15}_{-2.91}$&602$^{+35}_{-41}$&174$^{+90}_{-89}$&13.6$^{+5.1}_{-5.5}$&0.988$\pm{0.003}$ &407.84/417\\
&91015-01-03-03&240&0.01$^{+0.03}_{-0.03}$&0.98$^{+0.04}_{-0.04}$&31.0$^{+1.0}_{-0.9}$&-&-&-&2.58$^{+0.11}_{-0.12}$&3.96$^{+0.45}_{-0.38}$&17.6$^{+2.5}_{-2.2}$&31.4$^{+76.9}_{-136.3}$&435$^{+82}_{-80}$&38.0$^{+6.4}_{-6.2}$&1.100$\pm{0.004}$ &524.03/498\\
&91015-01-06-01&49&0.23$^{+1.41}_{-2.67}$&19.3$^{+4.0}_{-3.2}$&39.9$^{+6.2}_{-5.3}$&-&-&-&23.9$^{+1.3}_{-1.5}$&8.06$^{+6.21}_{-3.84}$&5.20$^{+3.83}_{-2.49}$&577$^{+39}_{-47}$&354$^{+193}_{-130}$&46.3$^{+14.8}_{-17.3}$&0.926$\pm{0.006}$ &449.82/420\\
&91015-01-06-02&25&2.62$^{+1.73}_{-2.87}$&25.9$^{+5.8}_{-4.6}$&29.2$^{+5.6}_{-4.7}$&-&-&-&fixed&211$^{+60}_{-43}$&44.5$^{+5.5}_{-6.2}$&681$^{+16}_{-15}$&115$^{+58}_{-38}$&16.3$^{+4.8}_{-4.1}$&0.863$\pm{0.004}$ &408.73/424\\
&91059-03-03-02&41&0.13$^{+0.06}_{-0.08}$&1.40$^{+0.13}_{-0.12}$&27.1$^{+1.7}_{-1.6}$&-&-&-&4.07$^{+0.27}_{-0.31}$&5.77$^{+0.91}_{-0.84}$&20.3$^{+2.7}_{-2.7}$&328$^{+18}_{-19}$&104$^{+56}_{-36}$&15.9$^{+5.8}_{-4.9}$&1.097$\pm{0.007}$ &446.60/421\\
&91057-01-04-01&789&0.01$^{+0.01}_{-0.02}$&0.69$^{+0.03}_{-0.02}$&22.6$^{+0.8}_{-0.7}$&-&-&-&1.25$^{+0.16}_{-0.16}$&4.80$^{+0.25}_{-0.24}$&22.4$\pm{1.6}$&141$^{+22}_{-23}$&300$^{+41}_{-39}$&17.3$\pm{1.7}$&1.009$\pm{0.006}$ &451.16/421\\
&92049-01-01-00&95&fixed&1.00$^{+0.06}_{-0.06}$&21.3$^{+1.0}_{-1.1}$&-&-&-&fixed&12.0$\pm{1.0}$&38.8$^{+1.5}_{-1.6}$&86.9$^{+49.7}_{-126.6}$&346$^{+141}_{-99}$&29.9$^{+6.8}_{-5.6}$&1.066$\pm{0.005}$ &455.33/425\\
&92049-01-44-00&25&25.0$\pm{0.9}$&7.01$^{+3.47}_{-2.40}$&3.76$^{+1.45}_{-1.13}$&-&-&-&-&-&-&916$^{+9}_{-8}$&54.0$^{+32.0}_{-20.1}$&8.40$^{+2.81}_{-2.29}$&0.705$\pm{0.003}$ &409.45/421\\
&93030-01-53-00&43&fixed&1.70$^{+0.12}_{-0.11}$&19.8$^{+1.6}_{-1.5}$&-&-&-&1.65$^{+1.29}_{-1.33}$&16.10$^{+0.93}_{-1.29}$&36.9$^{+2.4}_{-3.0}$&286$^{+37}_{-53}$&222$^{+138}_{-81}$&13.0$^{+4.5}_{-4.0}$&1.078$\pm{0.004}$ &291.21/322\\
&95030-01-40-00&237&fixed&0.89$^{+0.04}_{-0.04}$&19.8$^{+0.8}_{-0.7}$&-&-&-&fixed&8.69$^{+0.66}_{-0.66}$&35.5$^{+1.4}_{-1.8}$&125$^{+31}_{-44}$&240$^{+71}_{-56}$&19.6$^{+3.9}_{-3.6}$&1.101$\pm{0.006}$ &477.93/422\\
&94030-01-20-00&432&fixed&0.73$^{+0.03}_{-0.02}$&22.1$\pm{0.9}$&-&-&-&1.13$^{+0.32}_{-0.36}$&5.48$^{+0.74}_{-0.94}$&24.7$^{+5.0}_{-8.2}$&92.1$^{+108.8}_{-230.2}$&517$^{+145}_{-134}$&26.9$^{+8.2}_{-7.2}$&1.054$\pm{0.010}$ &454.84/421\\
&93030-01-64-00&276&fixed&1.65$^{+0.06}_{-0.06}$&19.5$\pm{0.7}$&-&-&-&3.95$^{+0.26}_{-0.27}$&7.15$^{+1.51}_{-1.35}$&15.8$^{+5.4}_{-5.2}$&10.7$^{+112}_{-193}$&524$^{+121}_{-92}$&35.2$\pm{7.2}$&1.118$\pm{0.007}$ &541.23/421\\ 
\hline
IGR\ J17480-2446&95437-01-07-00&581&-&22.7$\pm{0.9}$&3.28$^{+0.09}_{-0.09}$&47.1$^{+0.5}_{-0.6}$&16.3$^{+2.6}_{-2.3}$&0.92$^{+0.10}_{-0.09}$&-&-&-&838$^{+12}_{-7}$&55.4$^{+32.4}_{-24.8}$&0.45$^{+0.13}_{-0.11}$&0.620$\pm{0.001}$ &662.08/333\\
&95437-01-09-00&403&0.21$^{+1.75}_{-2.40}$&31.2$^{+2.9}_{-2.8}$&3.55$^{+0.21}_{-0.20}$&44.9$\pm{0.6}$&9.09$^{+2.83}_{-2.27}$&0.57$^{+0.13}_{-0.12}$&-&-&-&853$^{+4}_{-5}$&42.5$^{+13.2}_{-12.4}$&0.88$^{+0.17}_{-0.15}$&0.655$\pm{0.001}$ &382.28/330\\
&95437-01-10-01&364&4.88$^{+0.89}_{-1.82}$&16.2$^{+7.1}_{-4.4}$&2.45$^{+0.60}_{-0.51}$&34.8$^{+3.1}_{-5.0}$&26.4$^{+18.2}_{-11.3}$&1.49$^{+0.72}_{-0.61}$&-&-&-&871$^{+5.6}_{-8.0}$&36.4$^{+18.3}_{-18.1}$&1.19$^{+0.39}_{-0.35}$&0.673$\pm{0.001}$ &352.23/332\\
\hline
4U 1820-30&10075-01-01-01&331&25.2$\pm{0.6}$&22.1$^{+2.0}_{-1.9}$&4.21$^{+0.33}_{-0.33}$&-&-&-&-&-&-&893$\pm{15}$&168$^{+40}_{-33}$&2.70$^{+0.43}_{-0.40}$&0.832$\pm{0.001}$ &388.64/386\\
&10075-01-01-02&649&-&-&-&73.3$^{+4.3}_{-9.6}$&61.0$^{+37.8}_{-26.0}$&1.14$^{+0.52}_{-0.45}$&-&-&-&1072$\pm{4}$&67.4$^{+12.9}_{-10.7}$&1.41$^{+0.17}_{-0.16}$&0.785$\pm{0.001}$ &414.95/366\\
&20075-01-01-00&910&-&-&-&-&-&-&-&-&-&1057$\pm{4}$&65.2$^{+13.9}_{-12.2}$&0.81$^{+0.11}_{-0.10}$&0.771$\pm{0.000}$ &323.05/308\\
&20075-01-03-00&759&4.81$^{+0.75}_{-0.82}$&19.8$\pm{2.8}$&3.04$^{+0.60}_{-0.63}$&81.4$^{+1.6}_{-1.7}$&9.64$^{+4.49}_{-3.10}$&0.17$^{+0.05}_{-0.05}$&-&-&-&1130$^{+6}_{-9}$&48.5$\pm{37.5}$&0.40$^{+0.22}_{-0.14}$&0.767$\pm{0.000}$ &313.29/301\\
&20075-01-05-00&532&13.4$^{+0.7}_{-0.6}$&12.2$^{+2.4}_{-3.5}$&4.36$^{+2.09}_{-2.14}$&30.1$^{+0.5}_{-0.6}$&10.3$^{+2.0}_{-1.7}$&1.6$^{+0.4}_{-0.3}$&56$^{+1.9}_{-2.1}$&43.2$^{+12.6}_{-12.0}$&3.68$^{+1.65}_{-1.55}$&680$\pm{4}$&149$^{+14}_{-13}$&7.62$^{+0.54}_{-0.52}$&0.922$\pm{0.001}$ &294.21/304\\
&20075-01-10-01&555&-&-&-&70.65$^{+2.74}_{-3.23}$&43.6$^{+15.0}_{-13.0}$&1.04$^{+0.28}_{-0.38}$&-&-&-&1035$\pm{4}$&57.2$^{+10.0}_{-9.0}$&1.17$^{+0.14}_{-0.13}$&0.776$\pm{0.001}$ &308.17/307\\
&30053-03-01-00&996&30.5$^{+1.4}_{-1.5}$&20.5$^{+4.4}_{-5.1}$&1.23$^{+0.63}_{-0.55}$&75.1$^{+1.7}_{-1.6}$&27.7$^{+8.2}_{-8.0}$&0.67$^{+0.16}_{-0.19}$&-&-&-&1073$\pm{+2}$&44.3$^{+5.1}_{-4.9}$&1.41$^{+0.11}_{-0.10}$&0.772$\pm{0.000}$ &316.55/307\\
&30053-03-02-02&464&29.7$^{+1.1}_{-1.6}$&15.6$^{+4.4}_{-3.5}$&1.44$^{+0.77}_{-0.45}$&86.3$^{+4.6}_{-4.2}$&28.4$^{+15.0}_{-8.9}$&0.52$^{+0.19}_{-0.15}$&-&-&-&1084$\pm{3}$&48.6$^{+9.9}_{-7.7}$&1.45$^{+0.20}_{-0.18}$&0.774$\pm{0.001}$ &395.19/366\\
&30053-03-02-03&835&32.1$^{+1.4}_{-2.6}$&9.80$^{+9.00}_{-3.94}$&0.37$^{+0.50}_{-0.16}$&79.0$^{+2.7}_{-2.4}$&16.4$^{+9.1}_{-5.8}$&0.32$^{+0.14}_{-0.11}$&-&-&-&1059$^{+9.3}_{-11.2}$&134$^{+34}_{-23}$&1.91$^{+0.29}_{-0.26}$&0.784$\pm{0.001}$ &330.04/307\\
&30057-01-03-02&306&15.9$^{+0.9}_{-1.0}$&42.3$^{+3.3}_{-3.0}$&5.84$^{+0.24}_{-0.23}$&79.0$^{+2.4}_{-2.3}$&11.6$^{+11.2}_{-6.5}$&0.35$^{+0.19}_{-0.15}$&-&-&-&1064$^{+5}_{-4}$&39.2$^{+10.9}_{-10.8}$&1.27$^{+0.26}_{-0.24}$&0.785$\pm{0.001}$ &290.13/310\\
&30057-01-04-02&147&fixed&7.36$^{+5.76}_{-4.45}$&0.59$^{+0.28}_{-0.24}$&29.9$\pm{1.2}$&20.7$^{+3.5}_{-3.0}$&3.73$^{+0.41}_{-0.40}$&-&-&-&664$\pm{1}$&11.0$^{+4.6}_{-7.3}$&4.48$^{+3.82}_{-0.35}$&0.805$\pm{0.002}$ &391.84/392\\
&30057-01-06-00&376&6.90$^{+1.29}_{-1.57}$&46.3$^{+3.4}_{-3.1}$&4.51$^{+0.18}_{-0.18}$&69.8$^{+1.7}_{-1.6}$&15.5$^{+5.9}_{-4.3}$&0.42$^{+0.13}_{-0.11}$&-&-&-&1037$\pm{4}$&73.3$^{+12.6}_{-10.6}$&1.81$^{+0.20}_{-0.19}$&0.767$\pm{0.001}$ &306.82/310\\
&30057-01-06-04&99&12.5$^{+1.3}_{-1.5}$&42.2$^{+3.7}_{-3.2}$&6.56$^{+0.29}_{-0.28}$&83.0$^{+5.3}_{-6.0}$&13.4$^{+8.9}_{-10.7}$&0.34$^{+0.18}_{-0.22}$&-&-&-&1077$^{+1.5}_{-3.9}$&14.2$^{+17.3}_{-45.7}$&1.74$^{+26.63}_{-0.33}$&0.760$\pm{0.001}$ &295.72/318\\
&40017-01-18-00&818&8.45$^{+1.32}_{-1.59}$&47.6$^{+3.9}_{-3.6}$&4.05$^{+0.19}_{-0.19}$&65.6$^{+1.9}_{-1.7}$&14.0$^{+5.5}_{-4.2}$&0.35$^{+0.11}_{-0.10}$&-&-&-&1028$\pm{3}$&45.6$^{+9.2}_{-7.5}$&1.13$^{+0.16}_{-0.15}$&0.775$\pm{0.001}$ &371.75/367\\
&40017-01-20-00&138&-&-&-&75.3$^{+3.8}_{-6.1}$&38.9$^{+18.6}_{-12.1}$&0.94$^{+0.37}_{-0.26}$&-&-&-&1067$\pm{5}$&36.8$^{+7.6}_{-6.6}$&0.99$^{+0.17}_{-0.17}$&0.750$\pm{0.001}$ &319.00/315\\
&40017-01-24-00&496&0.79$^{+0.35}_{-0.52}$&11.48$^{+1.24}_{-0.94}$&19.64$^{+1.46}_{-1.14}$&13.15$^{+0.37}_{-0.38}$&5.70$^{+1.53}_{-1.49}$&2.81$^{+0.95}_{-0.88}$&22.93$^{+0.77}_{-0.95}$&18.46$^{+2.68}_{-2.09}$&9.33$^{+1.49}_{-1.17}$&464.94$^{+12.75}_{-13.04}$&192.15$^{+45.29}_{-32.81}$&10.12$^{+1.85}_{-1.54}$&1.020$\pm{0.002}$ &315.47/326\\
&40019-02-01-00&1304&31.1$^{+1.5}_{-1.9}$&26.8$^{+6.0}_{-6.4}$&2.44$^{+0.90}_{-0.93}$&67.2$^{+1.3}_{-3.7}$&4.43$\pm{10.81}$&0.13$^{+0.17}_{-0.06}$&-&-&-&1010$^{+7}_{-6}$&67.9$^{+20.2}_{-17.2}$&1.29$^{+0.25}_{-0.23}$&0.797$\pm{0.001}$ &375.96/312\\
&70030-03-04-00&212&1.04$^{+0.63}_{-0.81}$&16.3$^{+1.8}_{-1.6}$&17.9$^{+1.3}_{-1.2}$&17.5$^{+0.5}_{-0.5}$&6.96$^{+2.20}_{-1.59}$&3.00$^{+1.25}_{-0.86}$&28.9$^{+2.1}_{-2.2}$&24.9$^{+3.9}_{-3.7}$&7.51$^{+1.64}_{-1.44}$&527$\pm{11}$&181$^{+31}_{-28}$&10.9$^{+1.3}_{-1.4}$&1.003$\pm{0.002}$ &341.36/326\\
&70030-03-05-00&417&1.17$^{+0.48}_{-0.55}$&14.8$\pm{1.3}$&17.5$\pm{1.0}$&15.5$\pm{0.3}$&5.21$^{+1.43}_{-1.15}$&2.07$^{+0.70}_{-0.53}$&26.1$^{+1.1}_{-1.3}$&23.7$^{+3.0}_{-2.8}$&8.72$^{+1.34}_{-1.16}$&516$^{+11}_{-10}$&154$^{+26}_{-23}$&7.83$^{+0.98}_{-1.02}$&1.010$\pm{0.001}$ &364.99/326\\
&70030-03-05-01&403&1.09$^{+0.57}_{-0.83}$&19.1$^{+2.4}_{-1.9}$&19.6$^{+1.8}_{-1.5}$&16.2$\pm{0.6}$&5.36$^{+1.94}_{-1.65}$&1.22$^{+0.59}_{-0.45}$&28.5$^{+0.8}_{-0.9}$&22.8$^{+3.2}_{-2.9}$&6.9$\pm{1.2}$&541$\pm{8}$&162$^{+32}_{-27}$&7.93$^{+1.08}_{-1.03}$&1.009$\pm{0.001}$ &315.50/326\\
&70030-03-05-02&206&11.2$^{+1.4}_{-1.0}$&14.2$\pm{4.0}$&7.39$^{+2.30}_{-4.47}$&26.8$^{+1.1}_{-1.3}$&12.8$^{+4.0}_{-2.9}$&2.75$^{+1.22}_{-0.88}$&44.9$^{+1.1}_{-1.0}$&10.5$^{+4.9}_{-3.3}$&1.47$^{+0.71}_{-0.52}$&648$\pm{10}$&132$^{+28}_{-23}$&6.12$^{+0.91}_{-0.85}$&0.944$\pm{0.002}$ &313.18/324\\
&90027-01-03-05&190&14.9$\pm{0.6}$&13.8$^{+2.4}_{-2.3}$&6.15$^{+1.17}_{-1.30}$&32.6$^{+1.5}_{-1.6}$&17.6$^{+7.2}_{-5.2}$&2.66$^{+1.24}_{-1.02}$&69.4$^{+4.6}_{-7.2}$&57.4$^{+41.1}_{-21.2}$&4.99$^{+3.58}_{-2.33}$&732$\pm{10}$&157$^{+28}_{-24}$&7.18$^{+0.87}_{-0.83}$&0.898$\pm{0.002}$ &246.89/305\\
&90027-01-04-01&398&13.9$^{+1.3}_{-1.6}$&45.3$^{+5.1}_{-4.3}$&5.17$^{+0.30}_{-0.29}$&72.7$^{+3.4}_{-3.3}$&18.9$^{+9.6}_{-6.1}$&0.59$^{+0.23}_{-0.20}$&-&-&-&1057$\pm{5}$&56.5$^{+15.8}_{-12.6}$&1.82$^{+0.32}_{-0.30}$&0.793$\pm{0.001}$ &407.50/369\\
&90027-01-06-00&799&fixed&3.46$^{+0.12}_{-0.13}$&15.3$\pm{0.5}$&-&-&-&6.37$^{+0.18}_{-0.21}$&10.0$^{+0.8}_{-0.7}$&18.2$^{+1.6}_{-1.8}$&228$^{+15}_{-16}$&359$^{+36}_{-31}$&19.0$^{+1.3}_{-1.2}$&1.095$\pm{0.001}$ &380.54/325\\
&92030-02-04-00&103&6.93$^{+3.05}_{-4.94}$&40.4$^{+9.9}_{-7.6}$&3.81$^{+0.51}_{-0.44}$&62.0$^{+2.7}_{-2.3}$&12.8$^{+9.2}_{-5.4}$&0.74$^{+0.36}_{-0.29}$&-&-&-&960$^{+55}_{-50}$&166$^{+133}_{-84}$&1.85$^{+0.90}_{-0.79}$&0.793$\pm{0.002}$ &258.33/242\\
&94090-01-01-02&765&fixed&23.8$^{+1.7}_{-1.9}$&21.2$^{+1.3}_{-1.6}$&19.7$\pm{0.7}$&7.42$^{+4.04}_{-2.25}$&2.09$^{+1.29}_{-0.70}$&32.4$^{+1.1}_{-1.3}$&13.3$^{+5.0}_{-3.6}$&3.30$^{+1.10}_{-0.85}$&575$^{+13}_{-14}$&141$^{+33}_{-28}$&8.35$^{+1.52}_{-1.53}$&0.994$\pm{0.002}$ &314.43/327\\
\hline
\end{tabular}
}}}
\caption{Table with parameters of the multi-Lorentzian fits per source. We quote the centroid frequency ($\nu$), full width at half maximum (FWHM), rms-normalized integral power (IP) of L$_b$, L$_{LF}$, L$_h$ and L$_u$, hard color (HC) and $\chi^2/dof$. The full version, with all parameters for all sources, is available online.}
\label{tab:pars2}
\end{table}


\bsp	
\label{lastpage}
\end{document}